\documentclass[superscriptaddress, prd, aps,amsmath,amssymb,showpacs,showkeys, twocolumn]{revtex4-2}
\usepackage[dvips]{graphicx,color}
\usepackage{subfig}
\usepackage{times}
\usepackage{xcolor}
\usepackage[%
  colorlinks=true,
  urlcolor=blue,
  linkcolor=red,
  citecolor=blue
]{hyperref}
\usepackage{orcidlink}

\begin{document}
\title{Joule-Thomson expansion and Optical behaviour of Reissner-Nordstr\"om-Anti-de Sitter black holes in Rastall gravity surrounded by a Quintessence field}

\author{Dhruba Jyoti Gogoi\orcidlink{0000-0002-4776-8506}}
\email[Email: ]{moloydhruba@yahoo.in}

\affiliation{Department of Physics, Dibrugarh University,
Dibrugarh 786004, Assam, India.}

\author{Yassine Sekhmani \orcidlink{0000-0002-4325-5252}}
\email[Email: ]{sekhmaniyassine@gmail.com}

\affiliation{D\'epartement de Physique, Equipe des Sciences de la mati\'ere et du rayonnement, ESMaR, Facult\'e des Sciences, Universit\'e Mohammed V de Rabat, Rabat, Morocco.}

\author{Digbijay Kalita }
\email[Email: ]{dijay.kalita24@gmail.com}

\affiliation{Department of Physics, Dibrugarh University,
Dibrugarh 786004, Assam, India.}

\author{Naba Jyoti Gogoi\orcidlink{0000-0002-1693-9501}}
\email[Email: ]{gogoin799@gmail.com}

\affiliation{Department of Physics, Dibrugarh University,
Dibrugarh 786004, Assam, India.}

\author{Jyatsnasree Bora \orcidlink{0000-0001-9751-5614}}
\email[Email: ]{jyatnasree.borah@gmail.com}

\affiliation{Department of Physics, Dibrugarh University,
Dibrugarh 786004, Assam, India.}

\begin{abstract}
This paper deals with the thermodynamics, Joule-Thomson expansion and optical behaviour of a Reissner-Nordstr\"om-anti-de Sitter black hole in Rastall gravity surrounded by a quintessence field. The black hole solution obtained in this framework is different from a corresponding black hole in General Relativity. The black hole metric function, as well as the Hawking temperature, is affected by the presence of energy-momentum conservation violation. The presence of energy-momentum conservation violation also affects the isenthalpic and inversion temperature curves, and with an increase in the Rastall parameter, the inversion temperature rises slowly. The impacts of other parameters, such as charge, structural constant etc., are investigated and compared. The black hole shadow, as well as the energy emission rate of the black hole, decreases with an increase in the Rastall parameter. Hence, the black holes evaporate slowly in presence of energy-momentum conservation violation.
\end{abstract}


\maketitle
\section{Introduction}
The theory of General Relativity (GR) is one of the widely accepted and elegant theories of 
spacetime and gravity. Among the various predictions, the most 
enchanting prediction of GR is the black holes and the Gravitational Waves (GWs). About 
a hundred years of its prediction, the viability of GR got its strong footing after the 
direct detection of Gravitational Waves (GWs) by the twin detector of the Laser Interferometer 
Gravitational Wave Observatory (LIGO) \cite{2016_Abbott}. These ground-based detectors 
LIGO and Variability of Solar Irradiance and Gravity Oscillations (Virgo) detector systems 
predicted the binary black hole systems, which stood as a piece of valid evidence for the 
viability of GR. Furthermore, the direct detection of GWs paved a new path to test or 
constrain different alternative or modified theories of gravity, including GR. In the weak 
field regimes and also in the moderately relativistic regimes, GR is found to pass some 
experimental tests like in the solar system tests \cite{will2014}, and in binary pulsars 
\cite{hulse1975, damour1992}. Moreover, in the highly relativistic strong gravity regime 
connected to binary black holes, its validity has been established by the detection of the 
GWs by LIGO and Virgo detectors. Thus, these GWs have changed our understanding of the 
universe in a great way. The properties of these GWs may change depending on the type of 
different modified gravities. As in the case of metric f(R) gravity, GWs have three 
polarization modes: one is tensor plus or GR modes, another is tensor cross modes, and the 
third is scalar polarization mode. The tensor plus and cross modes propagate with the 
speed of light through spacetime and are, in general, transverse, traceless and massless 
in nature. Whereas the scalar polarization mode is a mixture of massless breathing mode 
and massive longitudinal mode \cite{Liang_2017, gogoi1, gogoi2}. The massless breathing 
mode is not traceless but transverse in nature. So, these new and interesting properties 
of GWs in different Modified Theories of Gravity (MTGs) propelled to study the properties 
of black holes (which are considered as the cleanest objects in the universe) and other 
compact stars like neutron stars (the potential candidates for the generation of GWs) in 
the of MTGs regime. From this important point, a plethora of scientists is involved in the 
study of their behaviours and properties in different MTGs.

Among the different MTGs proposed till now, the simplest of all is the f(R) theories 
gravity. Another widely studied and important MTG is the Rastall gravity 
\cite{1972_Rastall}. This gravity theory was originally introduced in 1972 by P. Rastall 
\cite{1972_Rastall}. At the time of its proposal, this theory was unable to draw 
significant attention from the scientific community. Recently, in the last few years, it has
drawn the attention of so many researchers. This theory is found to be interesting due 
to the presence of non-vanishing background curvature, and it violates normal 
conservation laws. In this particular gravity theory, by making the covariant divergence of
the energy-momentum tensor proportional to the covariant divergence of the Ricci curvature 
scalar, the original conservation law is modified. However, the usual conservation law can 
easily be retrieved by setting the background curvature to zero. Also, it is important to note
that the Rastall gravity is equivalent to GR in absence of any matter source. This MTG is 
recently used in different aspects. In this gravity theory, the black holes and neutron 
stars are studied earlier by various authors \cite{Oliveira,Heydarzade}. For the case of 
neutron stars, the static and spherically symmetric solutions are studied in Ref. 
\cite{Oliveira}. Again in this theory, studies on the black hole solutions are reported in 
Ref. \cite{Heydarzade}. In another article in the same year, a study on different black 
hole solutions is reported by considering surrounding perfect fluids \cite{Heydarzade2}. 
For the case of black holes surrounded by quintessence field in Rastall gravity, the 
quasinormal modes are studied in Ref. \cite{Liang}. In the same article, the authors 
have reported the behaviour of quasinormal modes with respect to the Rastall gravity model 
parameter. They have shown that when $\kappa \lambda < 0$, a more rapid damping is observed
for the gravitational, electromagnetic and massless scalar field. Furthermore, for $\kappa 
\lambda > 0$, they have reported slower damping of the gravitational field, 
electromagnetic field and massless scalar field. Also, for the previous case, the real 
frequencies of quasinormal modes were found to be larger in comparison to that in GR. 
Whereas they are smaller for the latter case. They have also inferred that the change of 
the real and imaginary quasinormal modes frequencies with $\kappa \lambda$ are similar for 
different values of $l$ and $n$. Another important Ref.s in this field are \cite{Xu, Lin, 
Hu}. For the case of normal black holes with non-linear electrodynamic sources, studies are 
reported in Ref. \cite{Balart}. This study was done in the GR regime. In two other recent
works in the Rastall gravity, the black hole solutions and quasinormal modes for black holes with non-linear electrodynamic sources have been reported in Refs. \cite{gogoi3, gogoi4}. These studies show that the properties of compact objects vary significantly in Rastall gravity depending on the type of surrounding field.

In a recent study, $P-V$ criticality and Joule-Thomson expansion of charged anti-de Sitter black holes in Rastall gravity have been extensively studied \cite{Meng2020}. In this study, the inversion curves and isenthalpic curves in the $T-P$ plane have been investigated, and the impacts of the Rastall parameter have been studied. However, in this study, there was no additional surrounding field apart from the cosmological constant. In another study, the Joule-Thomson expansion of Reissner-Nordstr\"om-anti-de Sitter black hole immersed in the perfect fluid dark matter has been studied recently \cite{Cao2021}. This study was done in GR, and it shows that the surrounding dark matter fluid has noticeable impacts on black hole thermodynamics as well as the Joule-Thomson expansion of the black hole. In presence of a cloud of string and quintessence, the Joule-Thomson expansion of a charged anti-de Sitter black hole in GR has been studied in Ref. \cite{Yin2021}. Apart from these, there are several recent studies dealing with Joule-Thomson expansion of black holes in different theories of gravity \cite{jt01, jt02, jt03, jt04, jt05}. Being motivated by these studies, in this work, we have considered the Joule-Thomson expansion of a charged anti-de Sitter black hole in the framework of Rastall gravity. In this paper, we shall consider a charged anti-de Sitter black hole surrounded by a quintessence field which has not been considered till now for the study of Joule-Thomson expansion to the best of our knowledge. Apart from the thermodynamical behaviour of the black hole, we have considered two optical behaviours of the black hole {\it viz.}, the shadow and the energy emission rate of the black hole, to have a more understanding of the black hole solution in Rastall gravity.

The paper is organised as follows. In the next section i.e. Section \ref{sec02}, we have discussed Rastall gravity in brief. In Section \ref{sec03}, we have obtained a charged anti-de Sitter black hole solution in the framework of Rastall gravity. A few characteristics of this black hole solution are studied in Section \ref{sec04}. In Section \ref{sec05}, we have studied the Joule Thomson expansion of the black hole. Section \ref{sec06} deals with the optical characteristics of the black hole. Finally, in Section \ref{sec07}, we have summarised our results with a brief conclusion.


\section{Rastall Gravity}\label{sec02}
This theory of gravity modifies the GR by violating the covariant conservation condition $T^{\mu \nu}_{\ ; \, \nu} = 0$. In this case, the conservation condition changes to a more generalized version as given below:
\begin{equation}\label{r1}
\nabla_\nu T^{\mu\nu} = a^\mu.  
\end{equation}
For the sake  of consistency of this gravity theory with Relativity, the right-hand side of \eqref{r1} must vanish when the background curvature or the scalar curvature becomes zero. To do so, one can conveniently set the four-vector $a^\mu$ as,
\begin{equation}\label{r2}
 a^\mu = \lambda \nabla^\mu R. 
\end{equation}
In this expression for the four vectors, the term $\lambda$ is a free parameter, and it is known as Rastall's parameter. Using these two equations (Eq.s \eqref{r1} and \eqref{r2}), the field equations for the Rastall gravity can be formulated as:
\begin{equation}
R_{\mu\nu}-\frac{1}{2}\left(\, 1 -2\beta\,\right) g_{\mu\nu}R=\kappa T_{\mu\nu}\ ,\label{E0}
\end{equation}
here the term $\beta = \kappa \lambda$ is a Rastall parameter $\lambda$ dependent parameter, and thus, for convenience, we shall term this as the Rastall parameter from the hereafter. Taking the trace of the above equation will result,
\begin{equation}
R=\frac{\kappa}{\left(4\,\beta-1\right)}\,T\ ,\quad \beta\ne 1/4. \label{E2}
\end{equation}

Now, in presence of cosmological constant $\Lambda$ the field equation can be expressed as
\begin{equation}\label{E1}
G_{\mu\nu}+ \Lambda g_{\mu \nu} + \beta g_{\mu\nu}  R  = \kappa T_{\mu\nu},
\end{equation}
where $G_{\mu\nu}$ is the standard Einstein tensor.


\section{Black hole solution in Rastall gravity surrounded by Quintessence}\label{sec03}

We first consider a spherically symmetric generic spacetime metric in Schwarzschild coordinates, as given below, in order to obtain the surrounded black hole solutions.
\begin{equation} \label{metric}
ds^2 = -f(r) dt^2 + \dfrac{dr^2}{f(r)} +r^2 d\Omega^2,
\end{equation}
in this metric, $f(r)$ is the metric function that depends on $r$ and $d\Omega^2 = d\theta^2 + \sin^2\theta d\phi^2$. Using the field Eq.~\eqref{E1}, we can now define the Rastall tensor as $\Theta_{\mu\nu} = G_{\mu\nu}+\Lambda g_{\mu\nu} + \kappa \lambda g_{\mu\nu} R$. The non-vanishing components of the field equation can be given as
\begin{eqnarray}\label{H}
&&{\Theta^{0}}_{0}=\frac{1}{r^2}\big[rf^{\prime}(r) + f(r) -1 \big]+ \Lambda+\beta R,\nonumber\\
&&{\Theta^{1}}_{1}=\frac{1}{r^2}\big[rf^{\prime}(r) + f(r) - 1  \big]+ \Lambda+\beta R,\nonumber\\
&&{\Theta^{2}}_{2}=\frac{1}{r^2}\big[rf^{\prime}(r)+\frac{1}{2}r^2 f^{\prime\prime}(r)\big]+ \Lambda+\beta R,\nonumber\\
&&{\Theta^{3}}_{3}=\frac{1}{r^2}\big[rf^{\prime}(r)+\frac{1}{2}r^2 f^{\prime\prime}(r)\big]+ \Lambda+\beta R,
\end{eqnarray}
here, the Ricci scalar has the following functional form
\begin{equation}\label{R}
R=-\frac { 1
}{{r}^{2}}\big[{r}^{2}f^{\prime\prime}(r) + 4r f^{\prime}(r) + 2\,f(r) -2 \big].
\end{equation}
As usual, in these expressions, the prime denotes the derivative with respect to the radial coordinate $r$. It is inferred from these expressions that the mixed Rastall tensor components are equal i.e. ${\Theta^{0}}_{0} = {\Theta^{1}}_{1}$ and ${\Theta^{2}}_{2} = {\Theta^{3}}_{3}$. This is due to the consequence of the spherical symmetric nature of the metric (\ref{metric}) in the mixed tensor form. In the context of the present work, we shall consider a general total stress-energy tensor $T^\mu_\nu$ defined by
\begin{equation} \label{gen_T}
{{T}^{\mu}}_{\nu}={E^{\mu}}_{\nu}+{\mathcal{T}^{\mu}}_{\nu},
\end{equation}
here ${E^{\mu}}_{\nu}$ represents the trace-free Maxwell tensor expressed as
\begin{equation}\label{E*}
E_{\mu\nu}={\frac{2}{\kappa}}\left(F_{\mu\alpha}{F_{\nu}}^{\alpha}-
\frac{1}{4}g_{\mu\nu}F^{\alpha\beta}F_{\alpha\beta}\right),
\end{equation}
and $F_{\mu\nu}$ is the antisymmetric Faraday tensor. This tensor satisfies the following conditions:
\begin{eqnarray}\label{max}
&&{F^{\mu\nu}}_{;\mu}=0,\nonumber\\
&&\partial_{[\sigma} F_{\mu\nu]}=0.
\end{eqnarray}
In presence of the spherical symmetry, these equations will result,
\begin{equation}
F^{01}=\frac{Q}{r^2},
\label{addeq}
\end{equation}
here the $Q$ parameter acts as the electrostatic charge in the theory. 
Thus, the Maxwell tensor can be represented as
\begin{equation}\label{E**}
{E^{\mu}}_{\nu}={\frac{Q^2}{\kappa r^4}}~\begin{pmatrix}-1 & 0 & 0 & 0 \\
0& -1 & 0 & 0 \\
0 & 0 & 1 & 0 \\
0 & 0 & 0 & 1\\
\end{pmatrix}.
\end{equation}

Furthermore, in Eq.~\eqref{gen_T}, the term ${\mathcal{T}^{\mu}}_{\nu}$ represents the energy-momentum tensor of the surrounding field, and it can be expressed as \cite{Kiselev}
\begin{eqnarray}\label{sur}
&&{\mathcal{T}^{0}}_{0}=-\rho_s(r),\nonumber\\
&&{\mathcal{T}^{i}}_{j}=-\rho_{s}(r)\alpha\left[-(1+3\bar{\beta})\frac{r_i r^j}{r_n r^n}+\bar{\beta}{\delta^{i}}_{j}\right],
\end{eqnarray}
where the $\alpha$ and $\bar{\beta}$ terms are related to the black hole internal structure of the surrounding field. The surrounding field's barotropic equations of state are,
\begin{equation}
p_s=\omega_s \rho_s, ~~~\omega_s=\frac{1}{3}\alpha,
\end{equation}
here the terms $p_s$ and $\omega_s$ represent the pressure and equation of state parameter, respectively. As discussed in the Refs.~\cite{Kiselev,gogoi3}, we can express the free parameter $\bar{\beta}$ as
 \begin{equation}
\bar{\beta}=-\frac{1+3\omega_s}{6\omega_s}.
\end{equation}
Eventually, the non-zero components of the ${\mathcal{T}^{\mu}}_{\nu}$ tensor are,
\begin{eqnarray}
&&{\mathcal{T}^{0}}_{0}={\mathcal{T}^{1}}_{1}=-\rho_s,\nonumber\\
&&{\mathcal{T}^{2}}_{2}={\mathcal{T}^{3}}_{3}=\frac{1}{2}(1+3\omega_s)\rho_s.
\end{eqnarray}

From the components of  Rastall field equations: ${\Theta^{0}}_{0}={T^{0}}_{0}$ and ${\Theta^{1}}_{1}={T^{1}}_{1}$ will result
 \begin{equation}\label{e00}
\frac{1}{r^2}\big(rf^{\prime} + f  -1 \big) + \Lambda-\frac {\beta}{{r}^{2}}\big({r}^{2}f^{\prime\prime} + 4r f^{\prime}+2\,f -2\big)=-\kappa\rho_s-\frac{Q^2}{ r^4},
\end{equation} 
and similarly ${\Theta^{2}}_{2}={T^{2}}_{2}$ and ${\Theta^{3}}_{3}={T^{3}}_{3}$ components give
\begin{align}\label{e22}
\frac{1}{r^2}\big(rf^{\prime}+\frac{1}{2}r^2 f^{\prime\prime}\big) &+ \Lambda-\frac {\beta
}{{r}^{2}}\big({r}^{2}f^{\prime\prime}  +4r f^{\prime} +2\,f  -2 \big)
\notag \\&=\frac{1}{2}(1+3\omega_{s} )\kappa\rho_{s}+\frac{Q^2}{ r^4}.
\end{align}
For the quintessence field using $\omega_s = -1/3$ and then solving Eq.s \eqref{e00} and \eqref{e22}, the general solution for the metric function can be obtained as
\begin{equation}\label{f1}
f(r) = 1 - \frac{2M}{r} +\frac{Q^2}{r^2} + N_s r^{\frac{4 \beta }{1-2 \beta }} - \frac{\Lambda  r^2}{3-12 \beta},
\end{equation}
and the energy density as given below
\begin{equation}\label{rho}
\rho_s (r)=\frac{(2 \beta +1) (4 \beta -1) N_s r^{\frac{4 \beta }{1-2 \beta }-2}}{(1-2 \beta )^2 \kappa }.
\end{equation}
Here $M$ and $N_s$  are the integration constants and they represent the black hole mass and the structural nature of the black hole surrounding the field, respectively. Again for the metric function $f(r)$ (Eq.~\eqref{f1}), the metric \eqref{metric} modifies as: 
\begin{widetext}
\begin{equation}\label{metric01}
ds^2=-\left(1 - \frac{2M}{r} +\frac{Q^2}{r^2} + N_s r^{\frac{4 \beta }{1-2 \beta }} - \frac{\Lambda  r^2}{3-12 \beta}\right)dt^2
+\frac{dr^2}{1 - \frac{2M}{r} +\frac{Q^2}{r^2} + N_s r^{\frac{4 \beta }{1-2 \beta }} - \frac{\Lambda  r^2}{3-12 \beta}}
+r^2 d\Omega^2.
\end{equation}
\end{widetext}

\section{Characteristics of the solution}\label{sec04}
In this section, we study a few characteristics of the black hole solution in Rastall gravity. Initially, we intend to study the singularity and uniqueness of the black hole, for which we calculate the corresponding scalars of the black hole metric. These are the Ricci scalar, the Ricci squared and the Kretschmann scalar.

The Ricci scalar for the above metric is given by
\begin{equation} \label{ricci_m01}
R = \frac{4 \Lambda }{1-4 \beta }-\frac{2 (2 \beta +1) N_s r^{\frac{4 \beta }{1-2 \beta }-2}}{(1-2 \beta )^2},
\end{equation}
the Ricci squared is given by
\begin{align} \label{riccisq_m01}
&R_{\mu\nu}R^{\mu\nu}= \\ \notag & \frac{4 (1-2 \beta )^4 {G_1}+4 \left(8 \beta ^2+2 \beta -1\right) (1-2 \beta )^2 G_2 N_s+{G_3} N_s^2}{(1-4 \beta )^2 (1-2 \beta )^4 r^8},
\end{align}
where $$G_1 = (1-4 \beta )^2 Q^4+\Lambda ^2 r^8,$$ $$G_2 = (1-4 \beta )^2 Q^2 r^{\frac{2}{1-2 \beta }}+\Lambda  r^{\frac{4 \beta }{1-2 \beta }+6}$$ and $$G_3 = 2 \left(8 \beta ^2-4 \beta +1\right) \left(8 \beta ^2+2 \beta -1\right)^2 r^{\frac{4}{1-2 \beta }}.$$
Finally, the Kretschmann scalar is found to be
\begin{align} \label{kret_m01}
R_{\alpha\beta\mu\nu}R^{\alpha\beta\mu\nu} =& \frac{8 \Lambda ^2}{3 (1-4 \beta )^2}+\frac{8 K_2 N_s r^{\frac{2}{1-2 \beta }-8}}{3 (1-2 \beta )^2 (4 \beta -1)} \notag \\ &+\frac{4 K_3 N_s^2 r^{\frac{4}{1-2 \beta }-8}}{(1-2 \beta )^4}+\frac{8 K_1}{r^8},
\end{align}
where
$$K_1 = 6 M^2 r^2-12 M Q^2 r+7 Q^4,$$
\begin{align}
    K_2 =& 3 (1-4 \beta )^2 \left(2 (1-6 \beta ) M r+(14 \beta -1) Q^2\right) \notag \\ &+(2 \beta +1) \Lambda  r^4 \notag
\end{align}
and
$$K_3 = 4 \beta  (\beta  (4 \beta  (14 \beta -9)+11)-2)+1.$$
From the expressions \eqref{ricci_m01}, \eqref{riccisq_m01} and 
\eqref{kret_m01}, it is seen that the black formed by this metric is singular 
for any value of the structural constant and the Rastall parameter. Basically, the singularity issue is arisen due to the mass term, charge term and the term with the structural parameter in the black hole metric. By choosing the Rastall parameter $\beta <\frac{1}{2}$, it is possible to remove the singularity issue arising from the term containing structural parameter $N_s$. However, to remove the singularity issue arising from the mass term and the charge term, one may introduce a non-linear charge distribution function similar to Ref. \cite{Balart}. In this study, we shall not consider such a situation and shall stick with the metric \eqref{metric01} for the rest of the analysis. Another observation made from the above scalars is that the Ricci scalar is not a function of the black hole charge $Q$. On the other hand, both the Ricci squared and the Kretschmann scalars are functions of the black hole charge $Q$, and any variation of the black hole charge can induce significant variations on the same. The scalars show that the black hole solution is unique, and the cosmological constant, as well as the surrounding field, changes the black hole spacetime significantly.

\begin{figure}[htb]
      	\centering{
      	\includegraphics[scale=0.40]{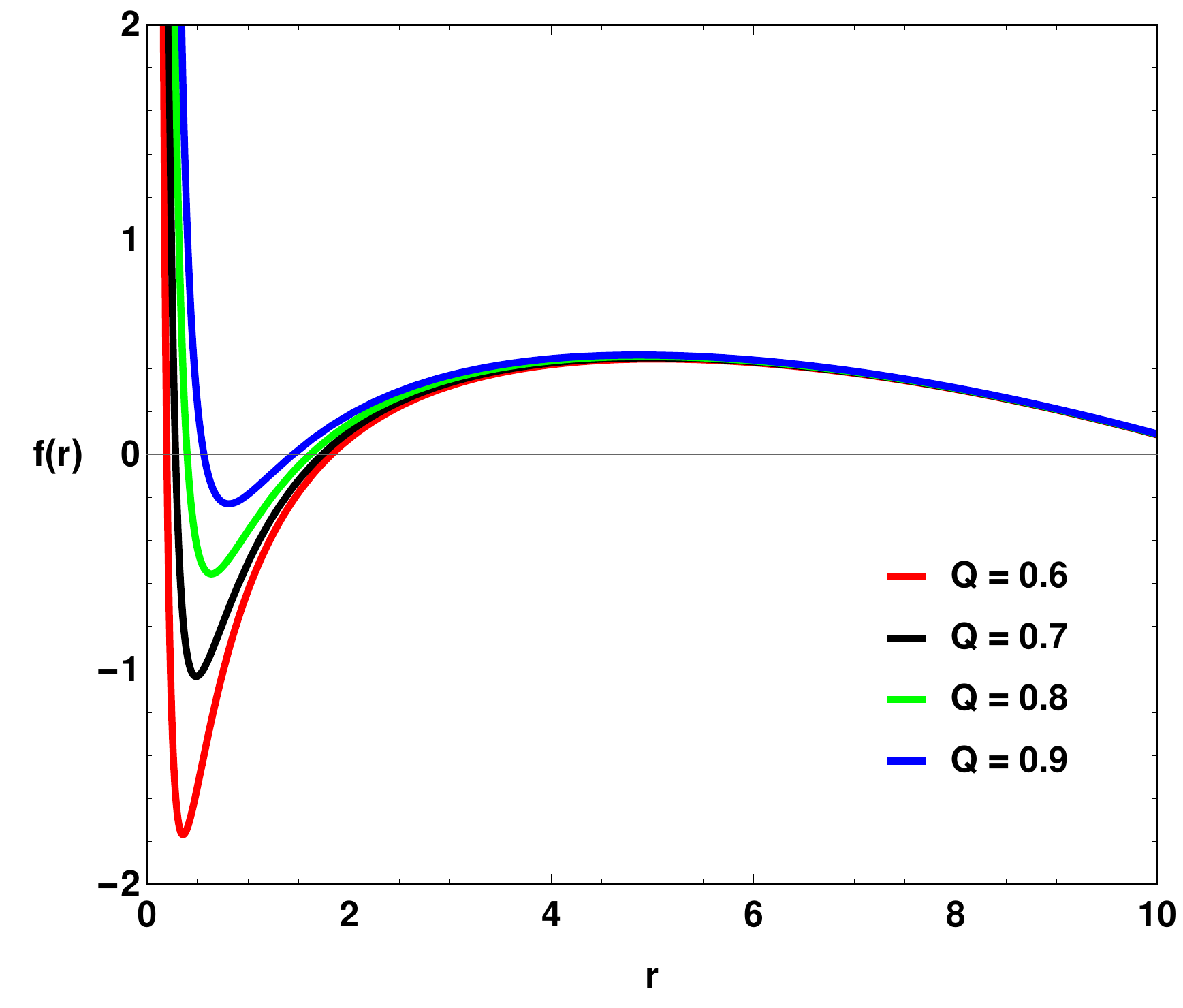}
       \includegraphics[scale=0.40]{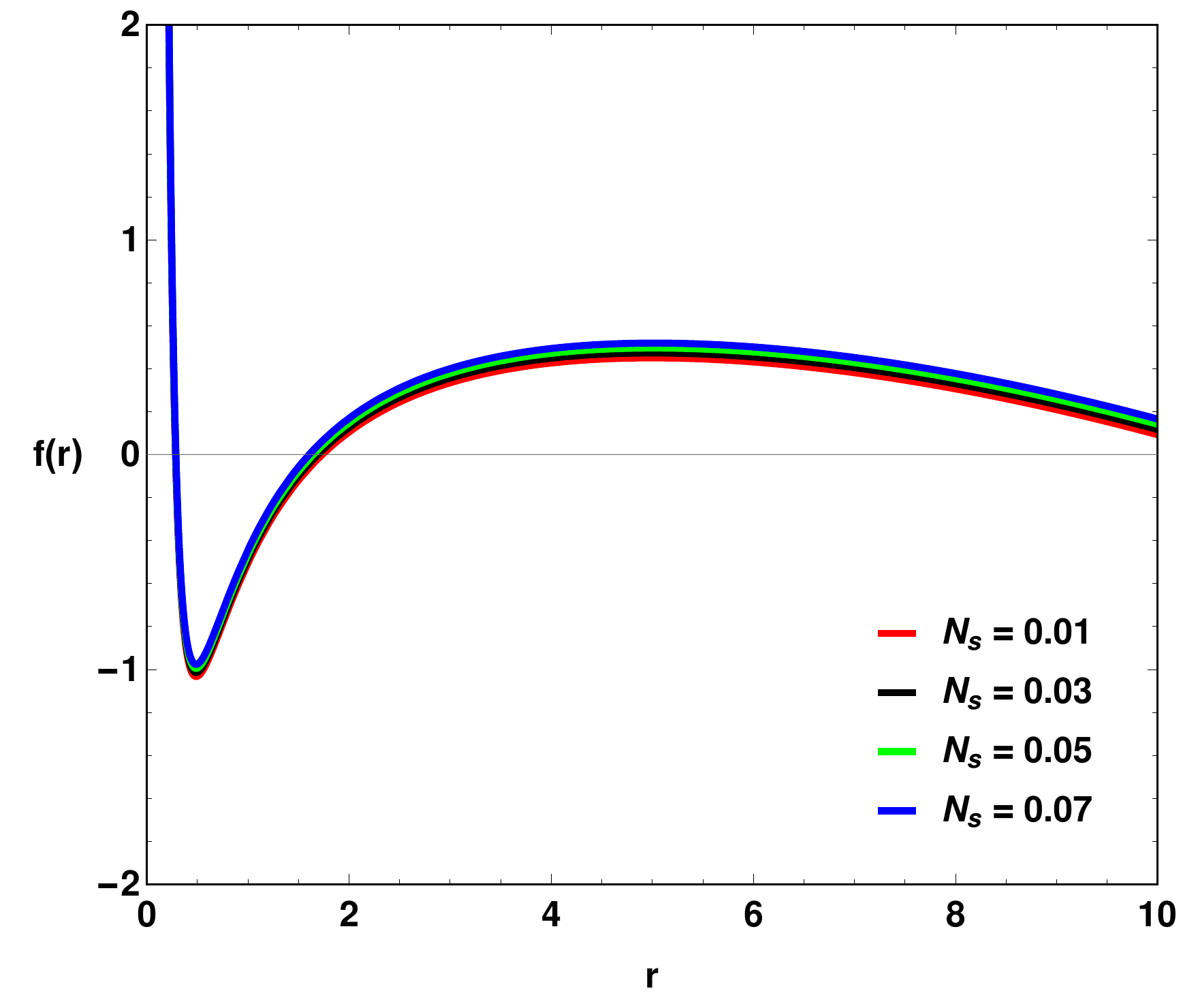}}
      	\caption{Variation of the black hole metric function \eqref{f1} with respect to $r$ with $M=1$, $\beta = 0.02$, $\Lambda = 0.02$ and on the first panel $N_s = 0.01$ and on the second panel $Q=0.7$.}
      	\label{figMetric01}
      \end{figure}
The horizons and the behaviour of the metric function of the black hole play an important role in this aspect. We use the metric function \eqref{f1} and study the preliminary characteristics of the black hole spacetime graphically. In Fig. \ref{figMetric01}, we plot the metric function for different values of black hole charge $Q$ and structural coefficient $N_s$. One can see that a variation in black hole charge has significant impacts over the Cauchy horizon and event horizon of the black hole. An increase in black hole charge reduces the separation between the Cauchy horizon and the event horizon. However, on the other hand, the black hole structural parameter $N_s$ does not have noticeable impacts over the Cauchy horizon. A decrease in this parameter increases the radius of the event horizon very slowly. But as one can see, this model parameter has noticeable impacts over the cosmological horizon of the black hole.
\begin{figure}[t]
      	\centering{
      	\includegraphics[scale=0.55]{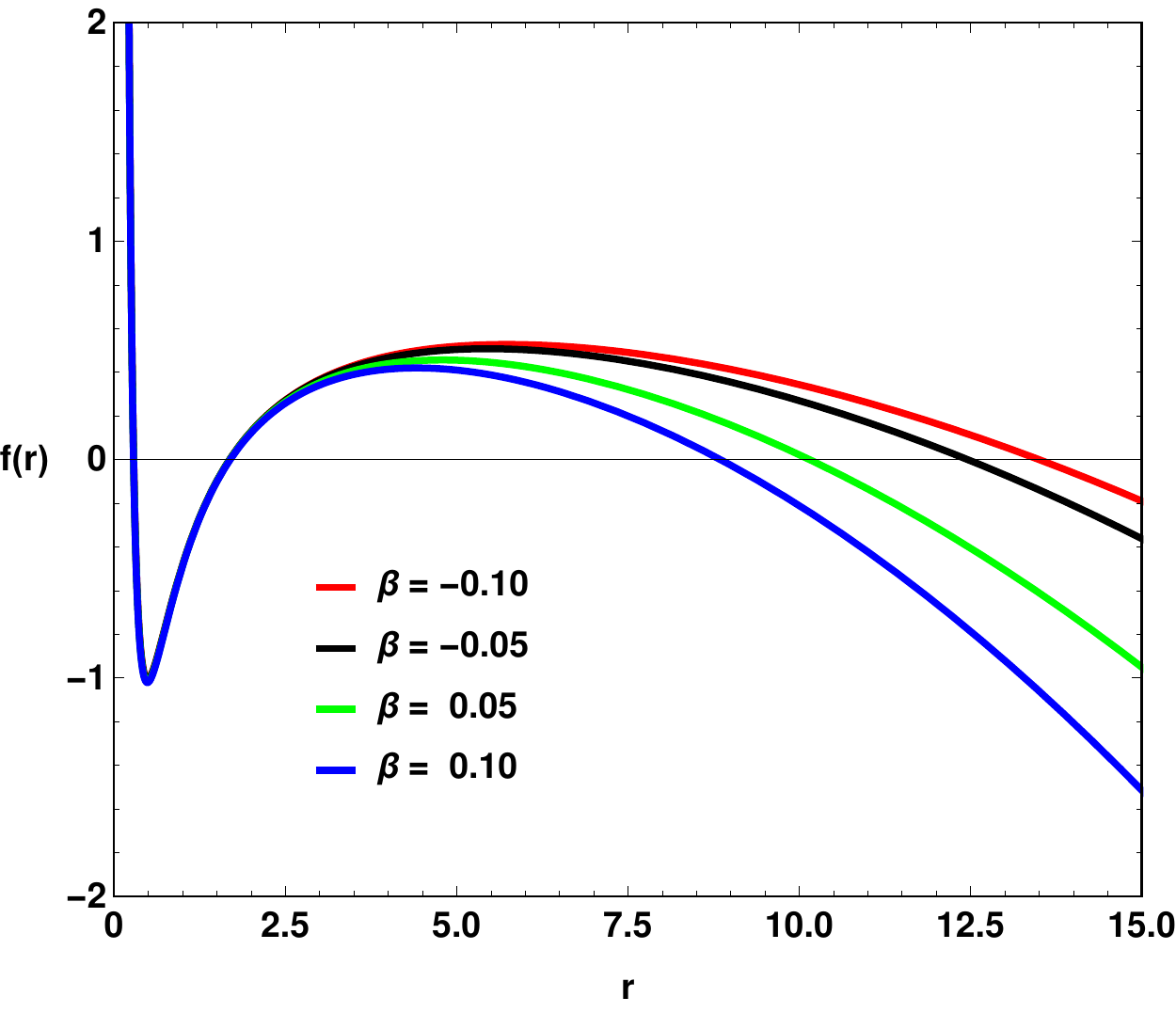}
       \includegraphics[scale=0.55]{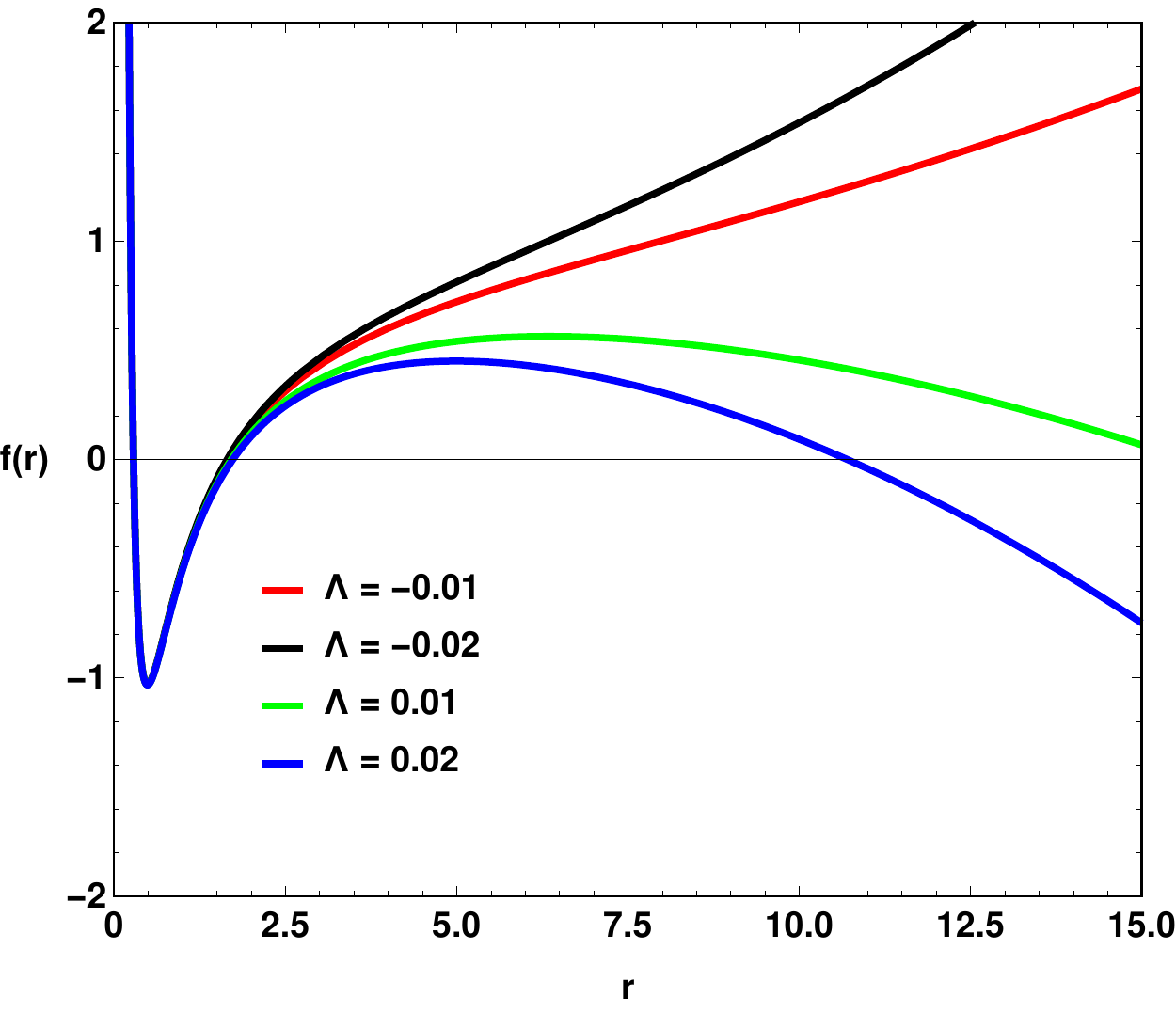}}
      	\caption{Variation of the black hole metric function \eqref{f1} with respect to $r$. On the first panel we have used $M=1$, $Q = 0.7$, $N_s = 0.03$ and $\Lambda = 0.02$ and on the second panel $M=1$, $Q = 0.7$, $N_s=0.01$ and $\beta=0.02$.}
      	\label{figMetric02}
      \end{figure}
In Fig. \ref{figMetric02}, we have shown the variation of the black hole metric function with respect to $r$ for different values of $\beta$ and $\Lambda$, respectively. On the first panel, we have seen that the Rastall parameter $\beta$ has impacts near the cosmological horizon of the black hole. In other words, the energy-momentum conservation violation has negligible impacts on the Cauchy horizon (first horizon) and the event horizon. With an increase in the parameter $\beta$, the cosmological horizon approaches the event horizon. On the second panel, we have shown the impacts of the cosmological constant on the black hole metric. Although the cosmological constant in $\Lambda CDM$ has a very small fixed value in general, in the framework of Rastall gravity, we have seen from the \eqref{metric01}, $\beta$ has an impact on the effective cosmological constant term. Hence, for the completeness of the study, we have studied the impacts of the cosmological constant also. The figure shows that in the case of the anti-de Sitter black hole, there is no cosmological horizon present. In this study, we shall consider the anti-de Sitter case only and hence, the black hole will have only two horizons {\it viz.}, the Cauchy horizon and the event horizon.

Here, we presume that the black hole pressure $P$ is associated with the cosmological constant present in the metric function \eqref{f1}. The associated pressure is given by
\begin{equation}
    P = - \frac{\Lambda}{8 \pi}.
\end{equation}
If $r_h$ is the event horizon radius of the black hole, it is possible to express the mass of the black hole in terms of the event horizon, as shown below:
\begin{equation}\label{mass_black_hole}
    M = -\frac{1}{2} r_h \left(-N_s r_h^{\frac{4 \beta }{1-2 \beta }}+\frac{8 \pi  P r_h^2}{12 \beta -3}-\frac{Q^2}{r_h^2}-1\right).
\end{equation}
The first law of black hole thermodynamics is given by
\begin{equation}
    dM = T dS + \Phi dQ + VdP + A d \beta,
\end{equation}
where $\Phi$ is electrostatic potential, and $A$ is a conjugate parameter associated with the violation of energy-momentum conservation violation.
The black hole entropy is associated with the area of the black hole, and hence, one can express black hole entropy as
\begin{equation}\label{entropy_black_hole}
    S = \pi  r_h^2.
\end{equation}
\begin{figure}[hbt!]
      	\centering{
      	\includegraphics[scale=0.55]{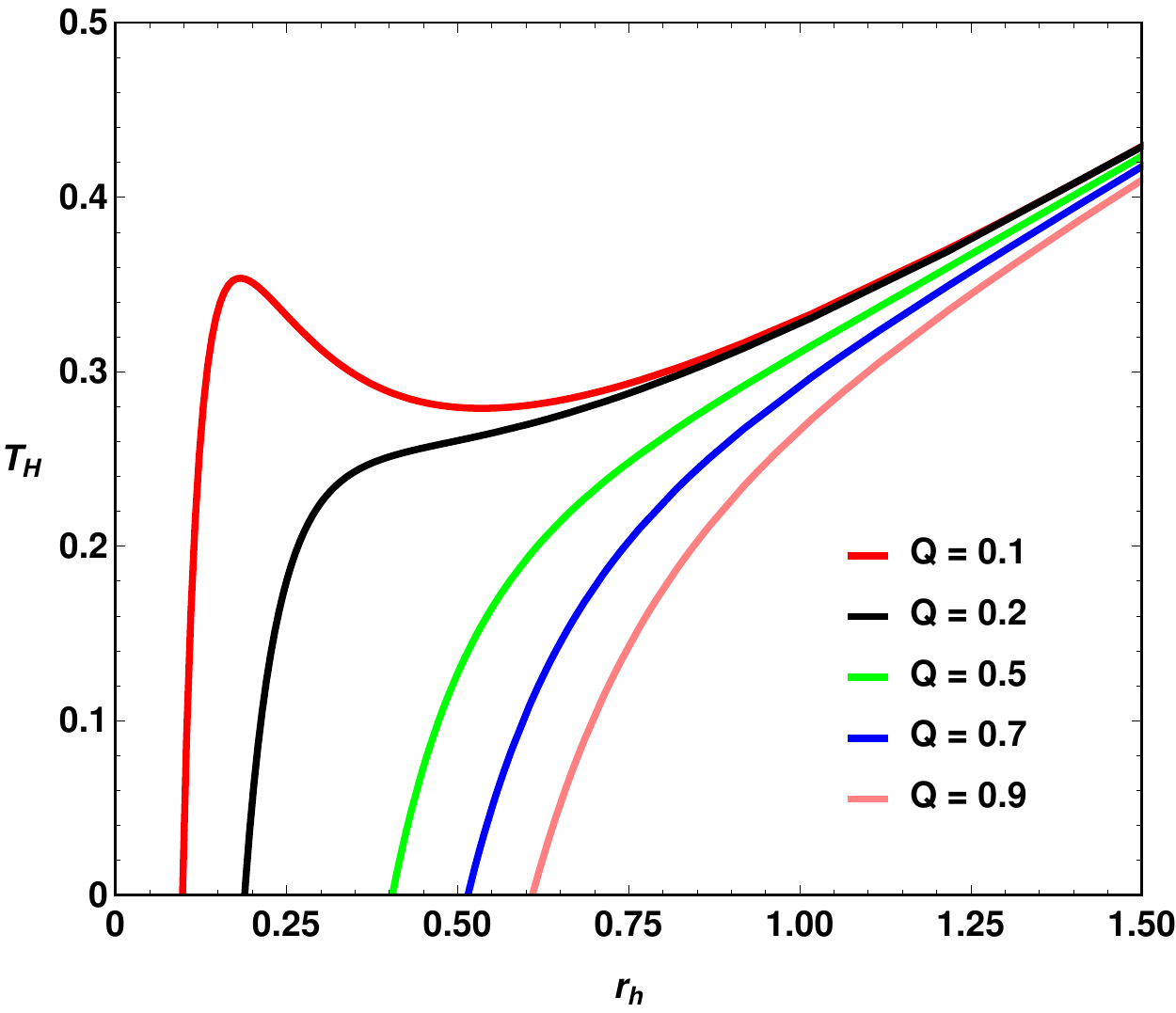}
       \includegraphics[scale=0.55]{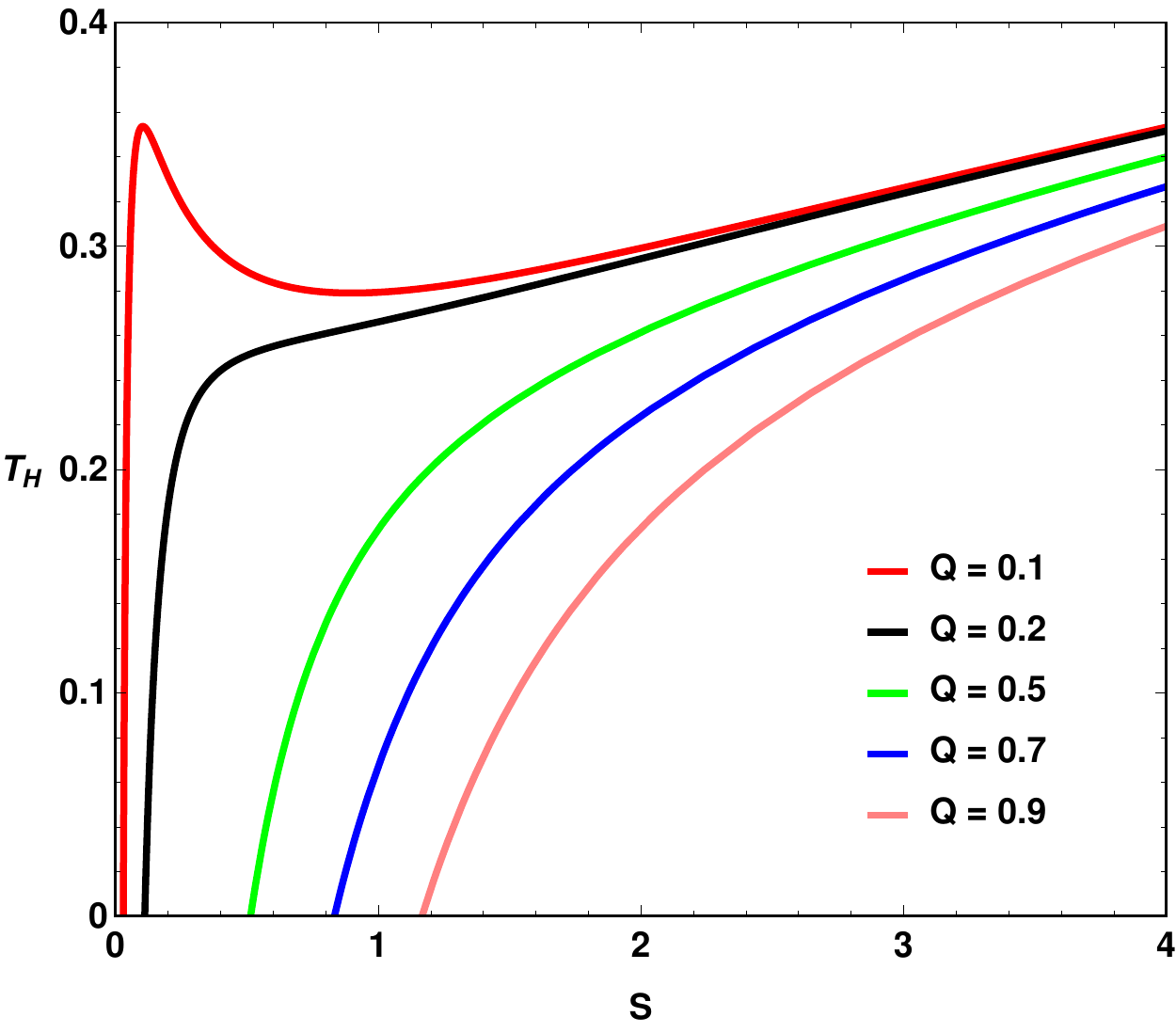}}
      	\caption{Variation of Hawking temperature $T_H$. We have used $N_s=0.01$, $P=0.075$, and $\beta=0.1$.}
      	\label{figTH01}
      \end{figure}
      
\begin{figure}[b!]
      	\centering{
      	\includegraphics[scale=0.55]{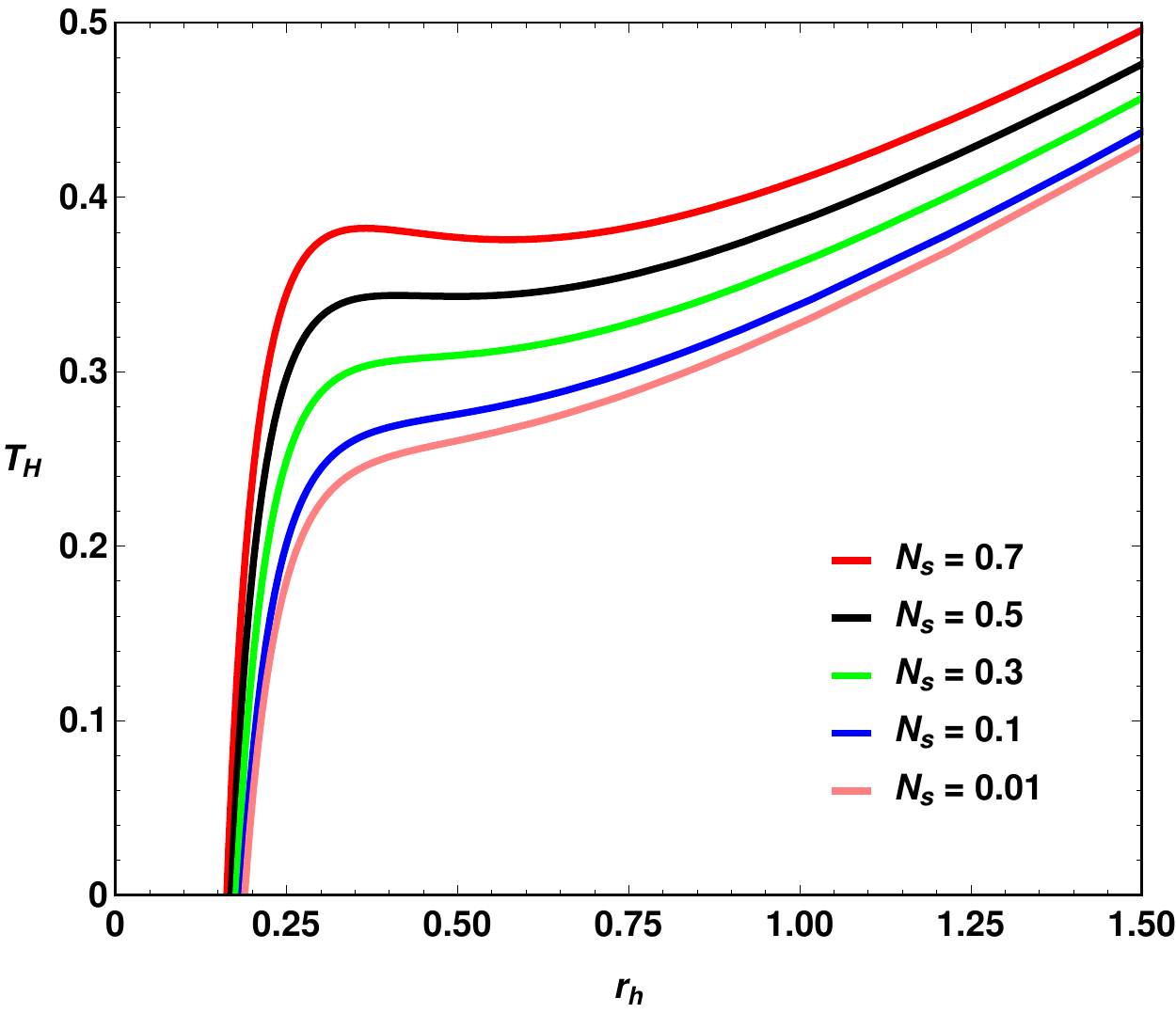}
       \includegraphics[scale=0.55]{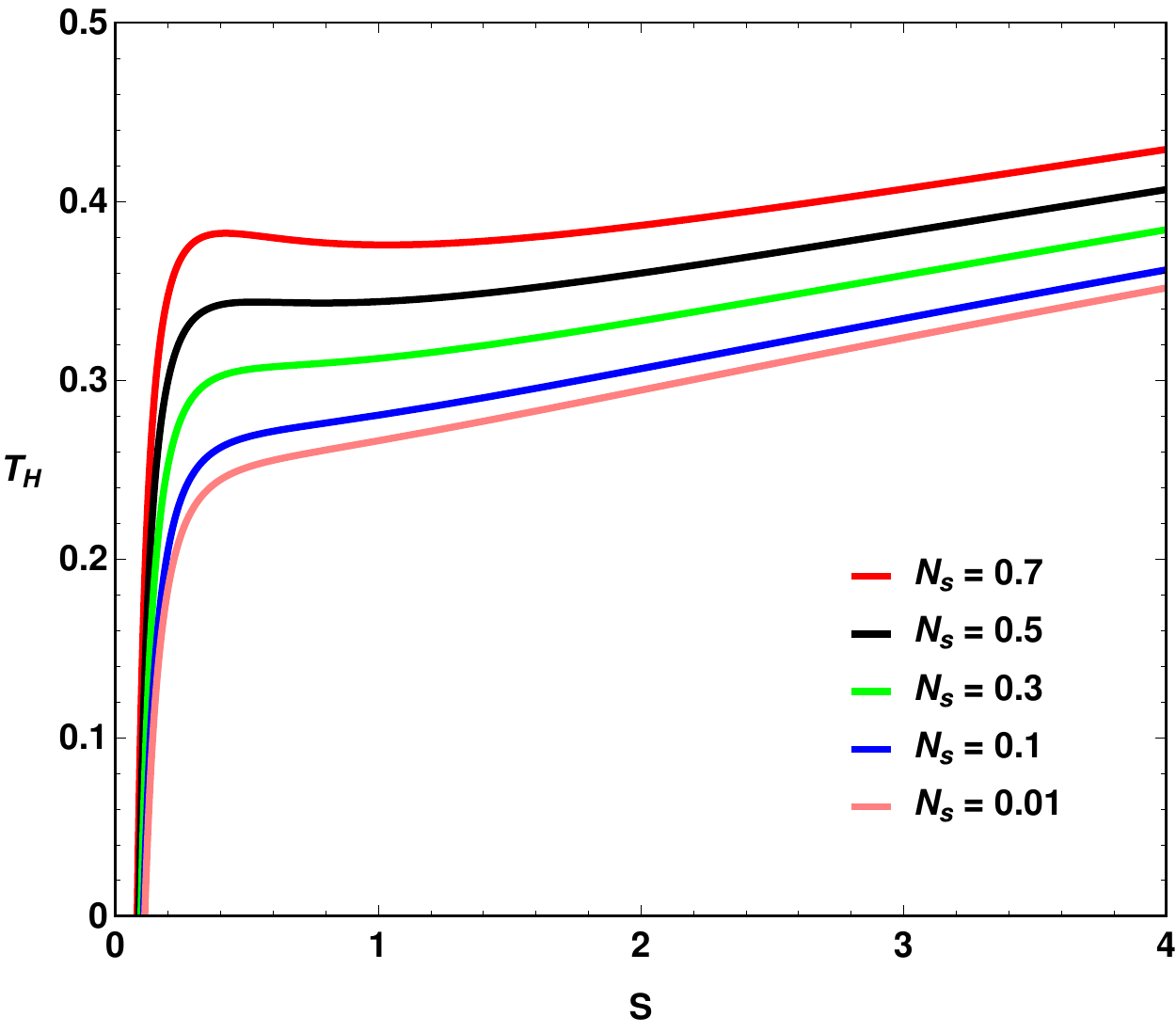}}
      	\caption{Variation of Hawking temperature $T_H$. We have used $Q=0.2$, $P=0.075$, and $\beta=0.1$. }
      	\label{figTH02}
      \end{figure}

\begin{figure}[h!]
      	\centering{
      	\includegraphics[scale=0.55]{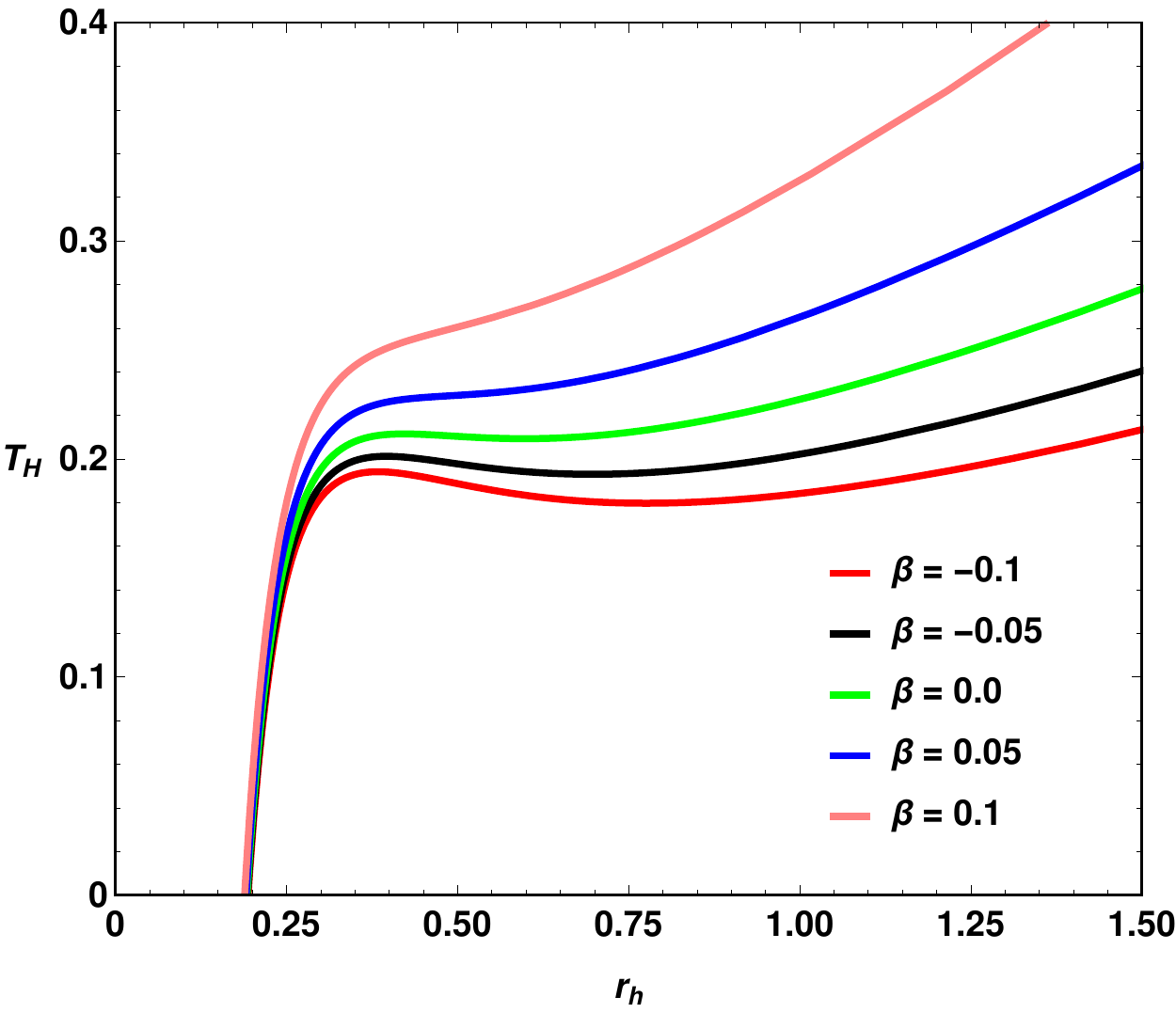}
       \includegraphics[scale=0.55]{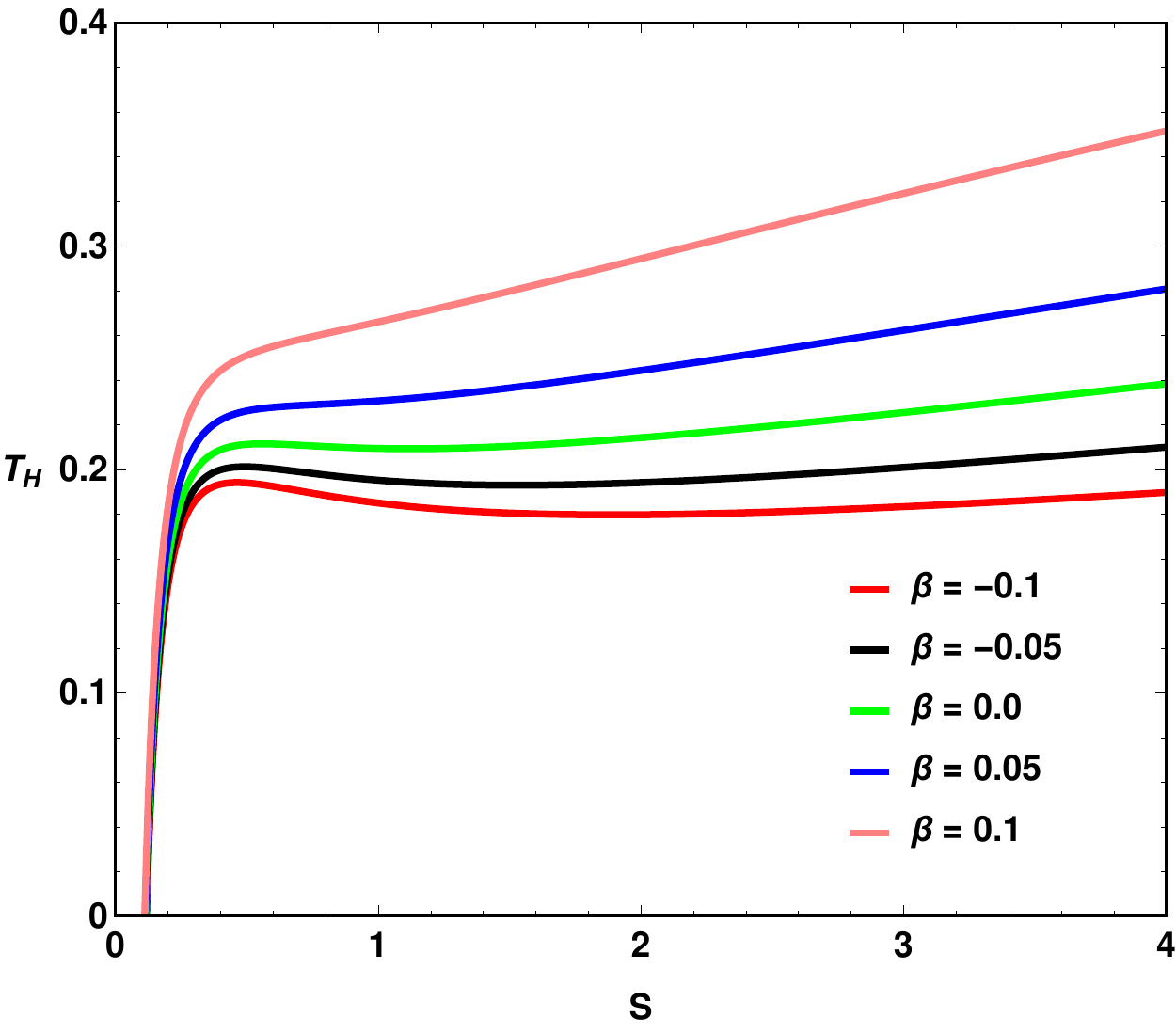}}
      	\caption{Variation of Hawking temperature $T_H$. We have used $Q=0.2$, $N_s=0.01$, and $P=0.075$. }
      	\label{figTH03}
      \end{figure}

 Using Eqs. \eqref{mass_black_hole} and \eqref{entropy_black_hole}, we calculate Hawking temperature of the black hole as
 \begin{widetext}
 \begin{align}\label{temperature}
     T_H &= \left(\frac{\partial M}{\partial S} \right)_{P, Q} \\ \notag 
     &= \frac{(2 \beta -1) \left\lbrace-8 \pi  P r_h^4+(4 \beta -1) r_h^2+(1-4 \beta ) Q^2\right\rbrace-(2 \beta +1) (4 \beta -1) N_s r_h^{\frac{2}{1-2 \beta }}}{4 \pi  \left(8 \beta ^2-6 \beta +1\right) r_h^3}\\ \notag
     &= \frac{-\pi ^{\frac{1}{2 \beta -1}+1} \left(8 \beta ^2+2 \beta -1\right) N_s S^{\frac{1}{1-2 \beta }}-(2 \beta -1) \left(S (-4 \beta +8 P S+1)+\pi  (4 \beta -1) Q^2\right)}{4 \sqrt{\pi } \left(8 \beta ^2-6 \beta +1\right) S^{3/2}}.
 \end{align}
\end{widetext}
We have graphically shown the variation of the Hawking temperature for different model parameters in Fig.s \ref{figTH01}, \ref{figTH02} and \ref{figTH03}. On the first panel of Fig. \ref{figTH01}, we have shown the variation of $T_H$ with respect to $r_h$ and on the second panel, variation with respect to $S$ respectively for different values of black hole charge $Q$. In both cases, we observe a peak or a crest of the curve for smaller values of the black hole charge $Q$. For higher values of charge, this crest decreases gradually. In Fig. \ref{figTH02}, we have considered different values of the parameter $N_s$. One can see that the parameter $N_s$ impacts $T_H$ in the opposite way in comparison to the case with different black hole charge $Q$. However, from Fig. \ref{figTH03}, one can see that with an increase in the parameter $\beta$, $T_H$ increases, and the crest is more distinct for smaller values of $\beta$.

 We use Eq. \eqref{temperature} to obtain the expression of pressure in terms of $T_H$ as
 \begin{widetext}
\begin{equation}
    P (r_h, T_H) =  -\frac{\left(8 \beta ^2+2 \beta -1\right) N_s r_h^{\frac{2}{1-2 \beta }}+\left(8 \beta ^2-6 \beta +1\right) \left(r_h^2 \left(4 \pi  T_H r_h-1\right)+Q^2\right)}{8 \pi  (2 \beta -1) r_h^4}.
\end{equation}
\end{widetext}
These expressions will be utilised to study the isenthalpic and inversion curves.

\section{Joule-Thomson expansion}\label{sec05}
In this section, we shall discuss the Joule-Thomson expansion of the black hole. One may note that the Joule-Thomson expansion exhibits the instability of a black hole system and the study of this behaviour, therefore, can shed some light on the property of the black hole. we calculate the Joule-Thomson coefficient $\mu$ of the black hole in order to determine its heating and cooling phases when dealing with the isenthalpic expansion. To obtain an expression of the Joule-Thomson expansion coefficient, we start with the specific heat at a constant pressure of the black hole, which can be obtained from the first law of thermodynamics as given by \cite{Meng2020}
\begin{equation}
    C_p = T \left( \frac{\partial S}{\partial T} \right)_{P,Q,\beta}.
\end{equation}
In this case, the isobaric heat capacity takes the following form:
\begin{equation}
    C_p = \frac{2 \pi  (\beta -1) \left[(\beta -1) r_h^{\frac{3 \beta }{\beta -1}} A_1- r_h^3 A_2\right]}{r_h \left[(\beta -1)^2 r_h^{\frac{3}{\beta -1}} \left(A_1+4 Q^2\right)+3 \beta  A_2\right]},
\end{equation}
where $A_1 = 8 \pi  P r_h^4-r_h^2-Q^2$ and $A_2 = (\beta +2) N_s.$ The fluids travel through a porous plug for the Joule-Thomson expansion of the van der Waals fluids from one side to the other, with pressure decreasing during the throttling process.
As a result, we use this analogous idea to understand black hole thermodynamics, and for that, we keep the black hole's enthalpy $M$ constant or fixed.
The Joule-Thomson coefficient $\mu$ is given by \cite{mu,Meng2020},
\begin{equation}
    \mu = \left( \frac{\partial T_H}{\partial P} \right)_M = \frac{1}{C_p} \left[ T_H \left( \frac{\partial V}{\partial T}\right)_P -V \right].
\end{equation}
During the expansion of a black hole, pressure always decreases, resulting in the change in pressure always being negative. On the other hand, a change in temperature in the process of expansion can be both positive and negative, depending on the situation. Hence, one can see that the sign of the Joule-Thomson coefficient $\mu$ depends on the sign of change in temperature during the process. When $\mu$ is positive, cooling occurs, and when $\mu$ is negative, heating takes place. 
   \begin{figure}[h!]
      	\centering{
      	\includegraphics[scale=0.5]{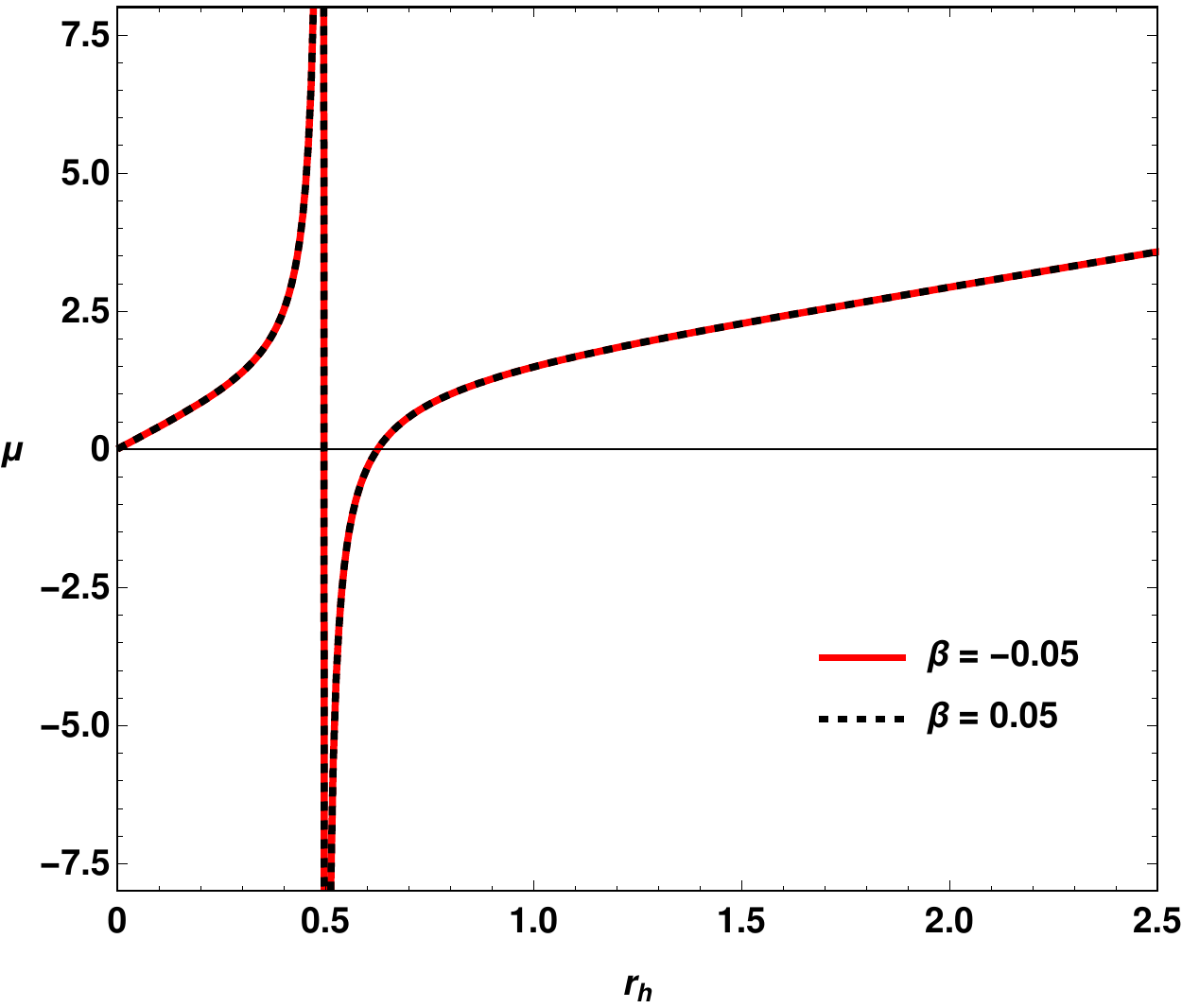}
       \includegraphics[scale=0.5]{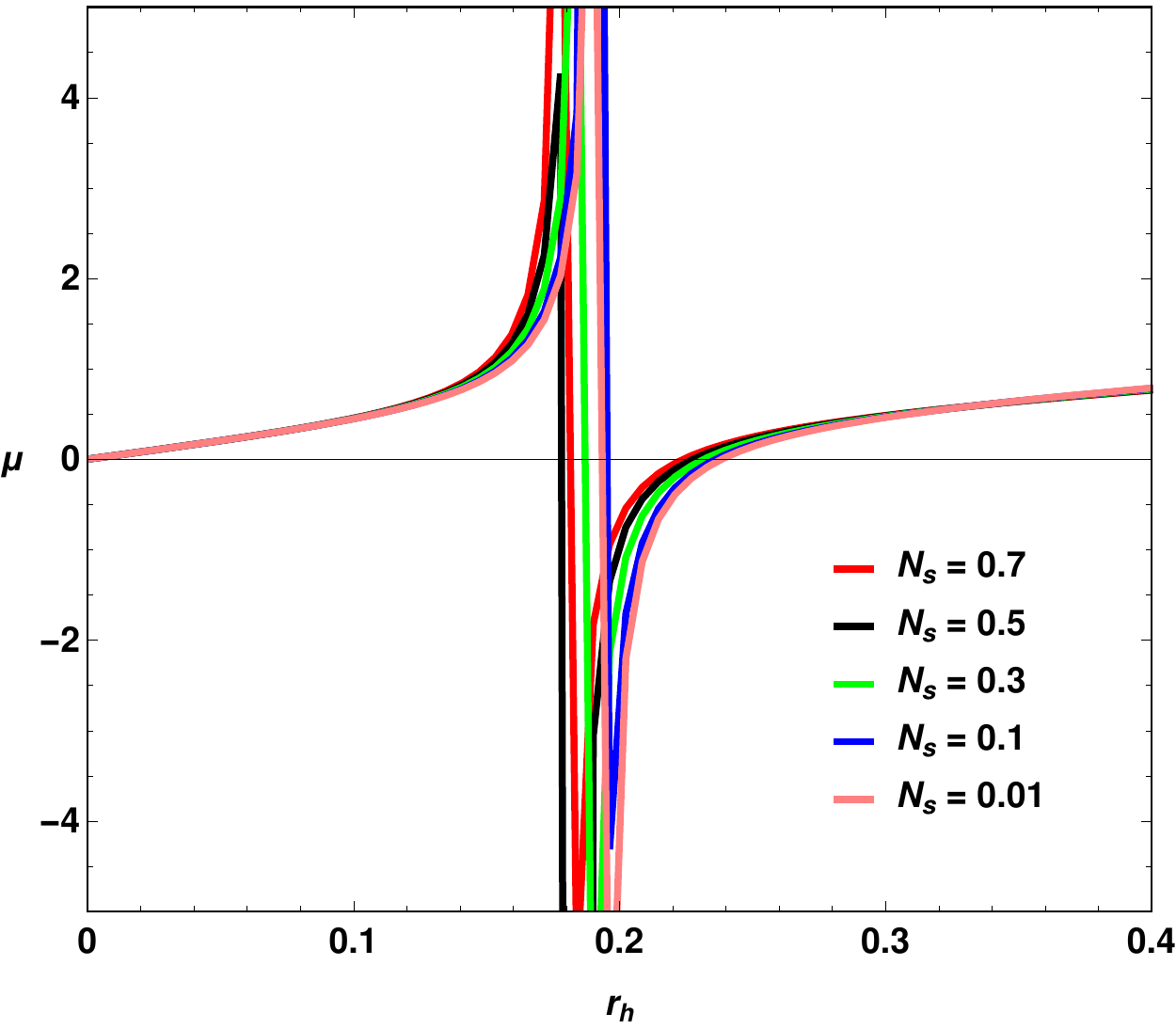}}
      	\caption{Variation of $\mu$.  We have used a) on the first panel $Q=0.6$, $N_s=0.01$, and $P=0.075$ and b) on the second panel $Q=0.2$, $\beta=0.1$, and $P=0.075$.  }
      	\label{figmu01}
      \end{figure}

    \begin{figure}[h!]
      	\centering{
      	\includegraphics[scale=0.5]{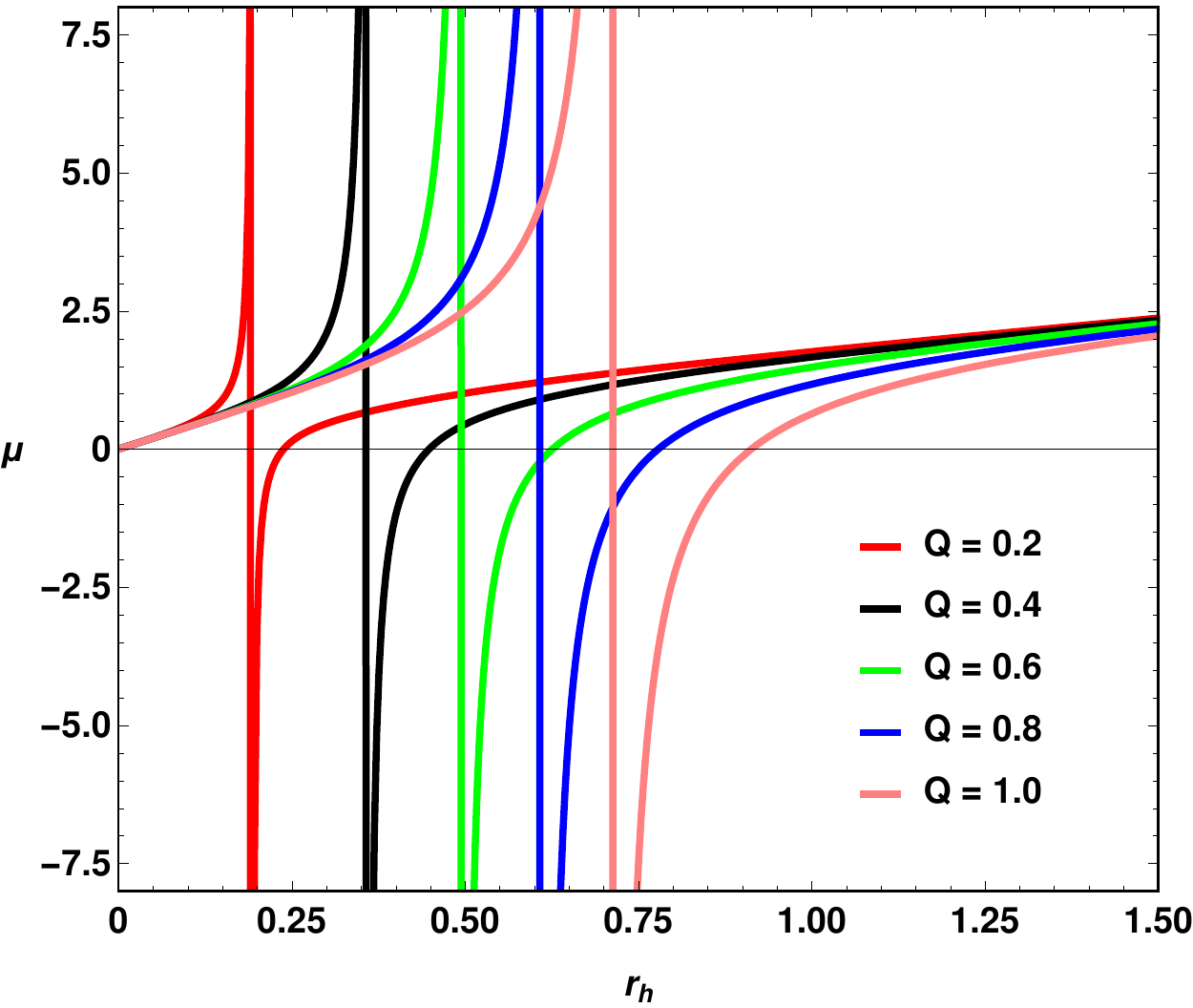}}
      	\caption{Variation of $\mu$. We have used $\beta=0.1$, $N_s=0.01$, and $P=0.075$. }
      	\label{figmu02}
      \end{figure}

In Fig.s \ref{figmu01} and \ref{figmu02}, we have shown the variation of the Joule-Thomson coefficient $\mu$ with respect to $r_h$ for different sets of model parameters. On the first panel of Fig. \ref{figmu01}, we have considered two different values of the Rastall parameter $\beta$, which depicts that the energy-momentum conservation violation has negligible effects on the Joule-Thomson coefficient $\mu$. On the second panel, we have considered different values of the black hole structural parameter $N_s$, which depicts that with an increase in the parameter, Joule-Thomson coefficient $\mu$ increases slowly and the inversion point shifts towards the smaller values of $r_h$. From Fig. \ref{figmu02}, one can see that black hole charge $Q$ has significant impacts on the Joule-Thomson coefficient $\mu$. For smaller values of charge $Q$, the inversion point shift towards smaller values of $r_h$ and when $Q$ increases, the coefficient $\mu$ increases gradually and the inversion point shifts towards higher values of $r_h$. This analysis indicates that the Joule-Thomson coefficient is highly impacted by the charge of the black hole while the structural parameter $N_s$ has comparatively smaller impacts on it. On the contrary, the Rastall parameter $\beta$, which accounts for the measurement of energy-momentum conservation violation in the theory, has almost negligible impacts on $\mu$.

\begin{figure}[h]
      	\centering{
      	\includegraphics[scale=0.50]{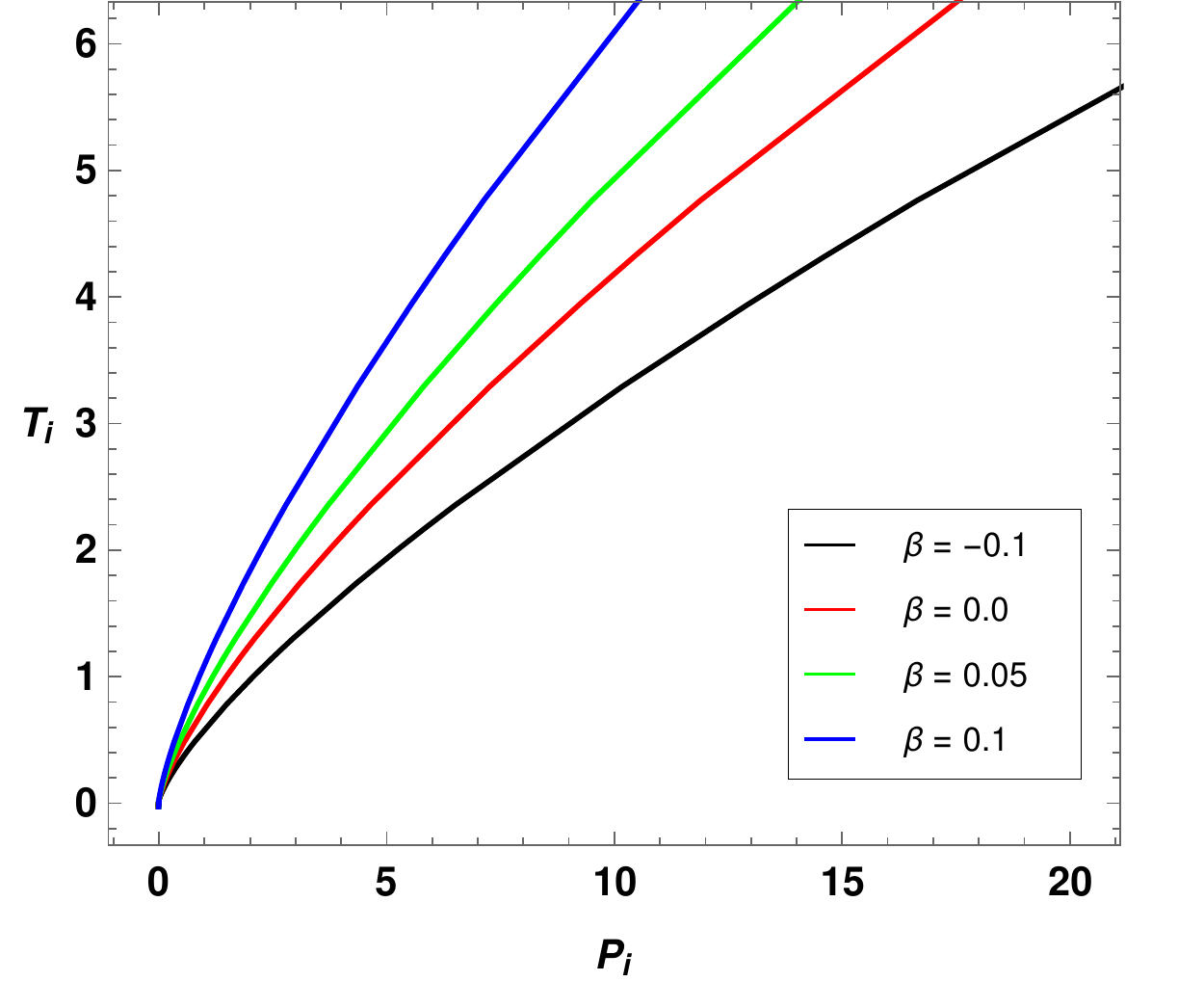}
       \includegraphics[scale=0.50]{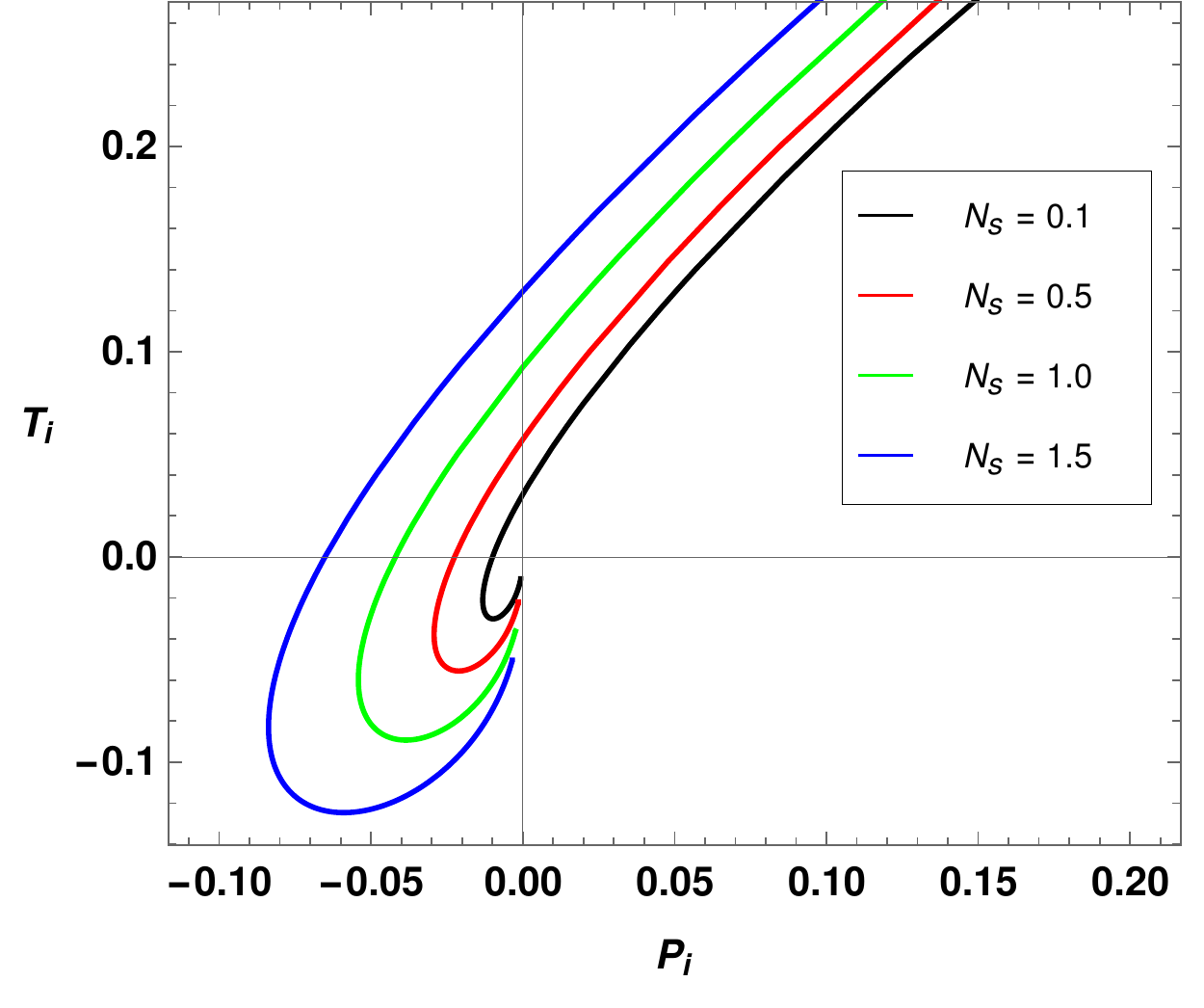}
      \includegraphics[scale=0.50]{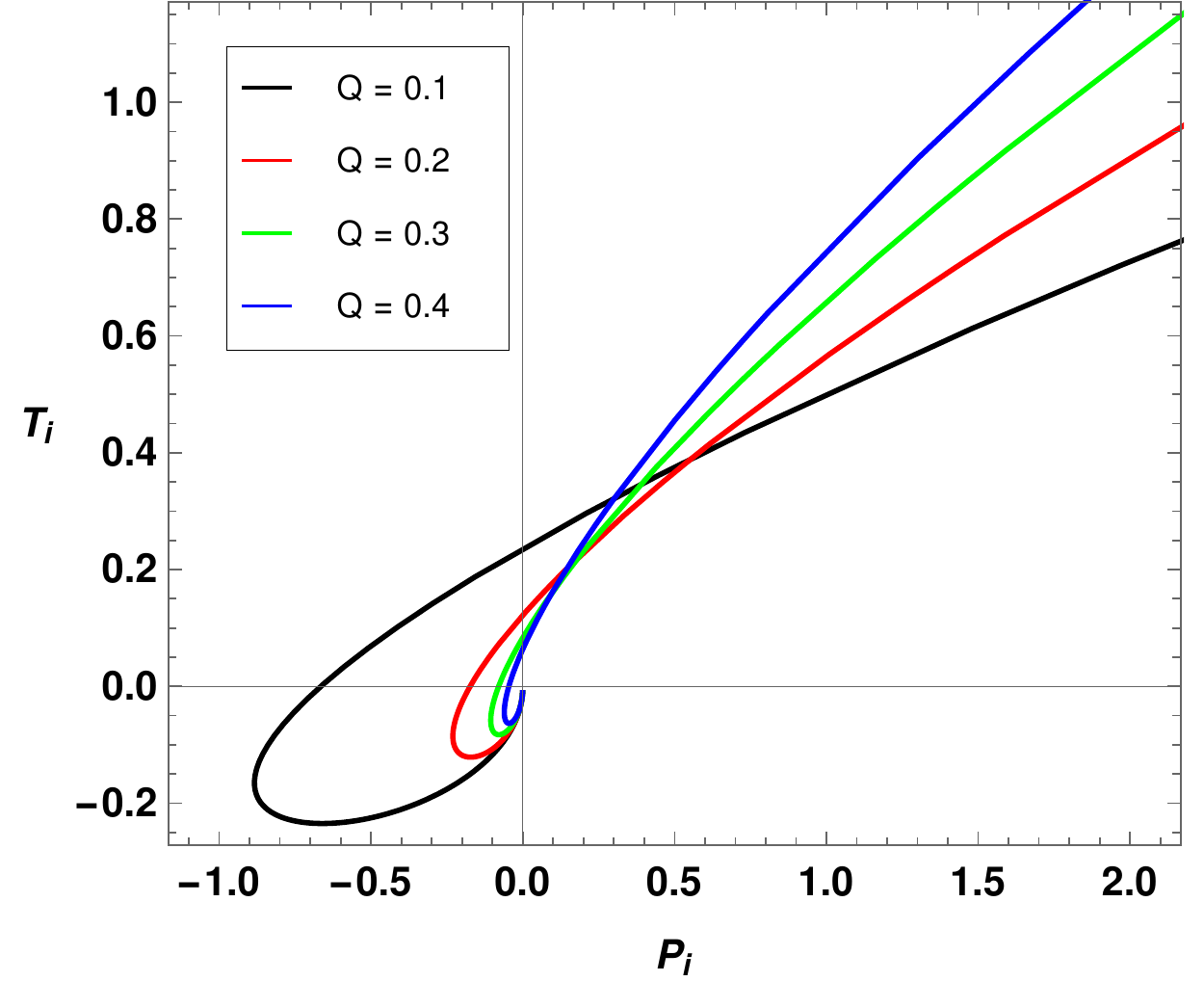}
       }
      	\caption{Variation of the inversion curves of the black hole defined by Eq. \eqref{metric01}. Here, we have considered in a) panel 01: $N_s = 0.05$ and $Q=0.9$, b) panel 02: $\beta = 0.1$ and $Q=0.9$ and c) panel 03: $N_s = 0.1$ and $\beta = 0.1$. }
      	\label{fig_invT_01}
      \end{figure}

The interesting and important behaviour of $\mu$ is basically the inversion point, and the temperature associated with the inversion point is denoted by the term inversion temperature. The inversion temperature associated with the black hole can be obtained by simply setting the Joule-Thomson coefficient $\mu=0,$ which gives,
\begin{equation}
    T_i = V \left( \frac{\partial T_H}{\partial V} \right)_P.
\end{equation}
For the black hole defined by the metric \eqref{metric01}, we found the inversion temperature as
\begin{equation}\label{tinvp}
    T_i = \frac{\frac{(2 \beta +1) (6 \beta -1) N_s r_h^{\frac{2}{1-2 \beta }}}{(1-2 \beta )^2}+\frac{8 \pi  P_i r_h^4}{1-4 \beta }-r_h^2+3 Q^2}{12 \pi  r_h^3},
\end{equation}
where $P_i$ is the pressure corresponding to the inversion temperature.
One can find the pressure corresponding to the inversion temperature as
\begin{equation}
    P_i = \frac{(4 \beta -1) \left(p_0-p_1 N_s r_h^{\frac{2}{1-2 \beta }}\right)}{8 \pi  (1-2 \beta )^2 r_h^4},
\end{equation}
where we have defined variable $p_0 = (1-2 \beta )^2 \left(2 r_h^2-3 Q^2\right)$ and the other variable $p_1 = 2 \left(6 \beta ^2+\beta -1\right).$
We can use this expression in Eq. \eqref{tinvp} to obtain the final expression of inversion temperature by eliminating the pressure term $P_i.$ The expression for $T_i$ is found to be
\begin{equation}
    T_i = \frac{\frac{\left(8 \beta ^2+2 \beta -1\right) N_s r_h^{\frac{2}{1-2 \beta }}}{(1-2 \beta )^2}-r_h^2+2 Q^2}{4 \pi  r_h^3}.
\end{equation}
We have shown the variation of the inversion temperature $T_i$ with respect to the corresponding pressure $P_i$ in Fig. \ref{fig_invT_01} for different values of the model parameters. On the first panel, one can see the variation of $T_i$ with respect to $P_i$ for different values of Rastall parameter $\beta$. One can see that with an increase in the Rastall parameter $\beta$, the inversion temperature of the black hole increases for higher pressure $P_i$. On the second panel, we have plotted the inversion temperature with respect to the pressure $P_i$ for different values of the black hole structural parameter $N_s$ and we observe that with an increase in the parameter, both the inversion temperature and pressure become negative initially. For smaller values of the parameter $N_s$, we can see that the negative portion of the inversion curve decreases. On the third panel of Fig. \ref{fig_invT_01}, we have shown the inversion curves for different values of the black hole charge $Q$. One can see that the impact of the black hole charge on the inversion curves is significantly different in comparison to the previous situations for different values of $\beta$ and $N_s$.

\begin{figure}[h]
      	\centering{
      	\includegraphics[scale=0.55]{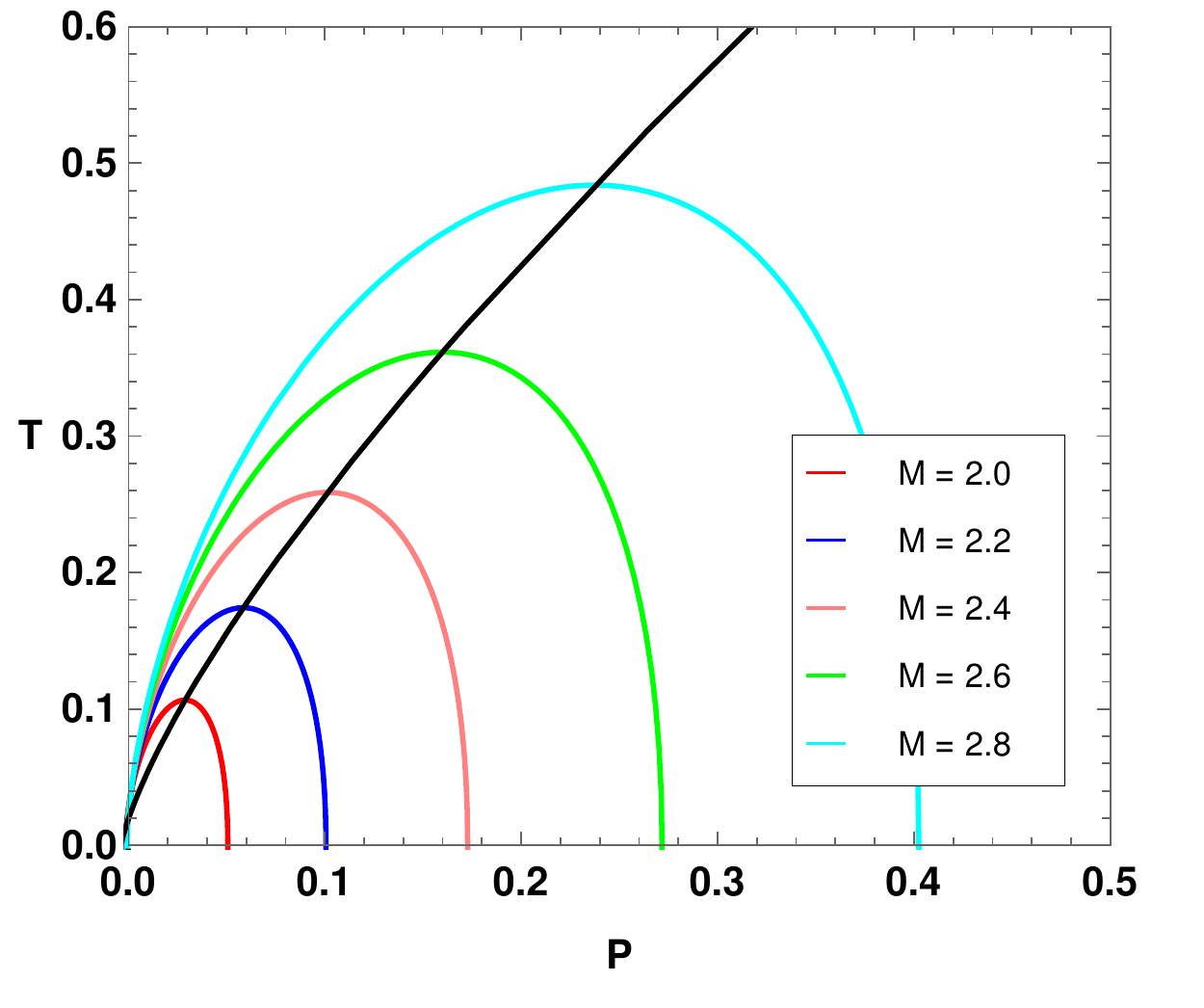}
       \includegraphics[scale=0.55]{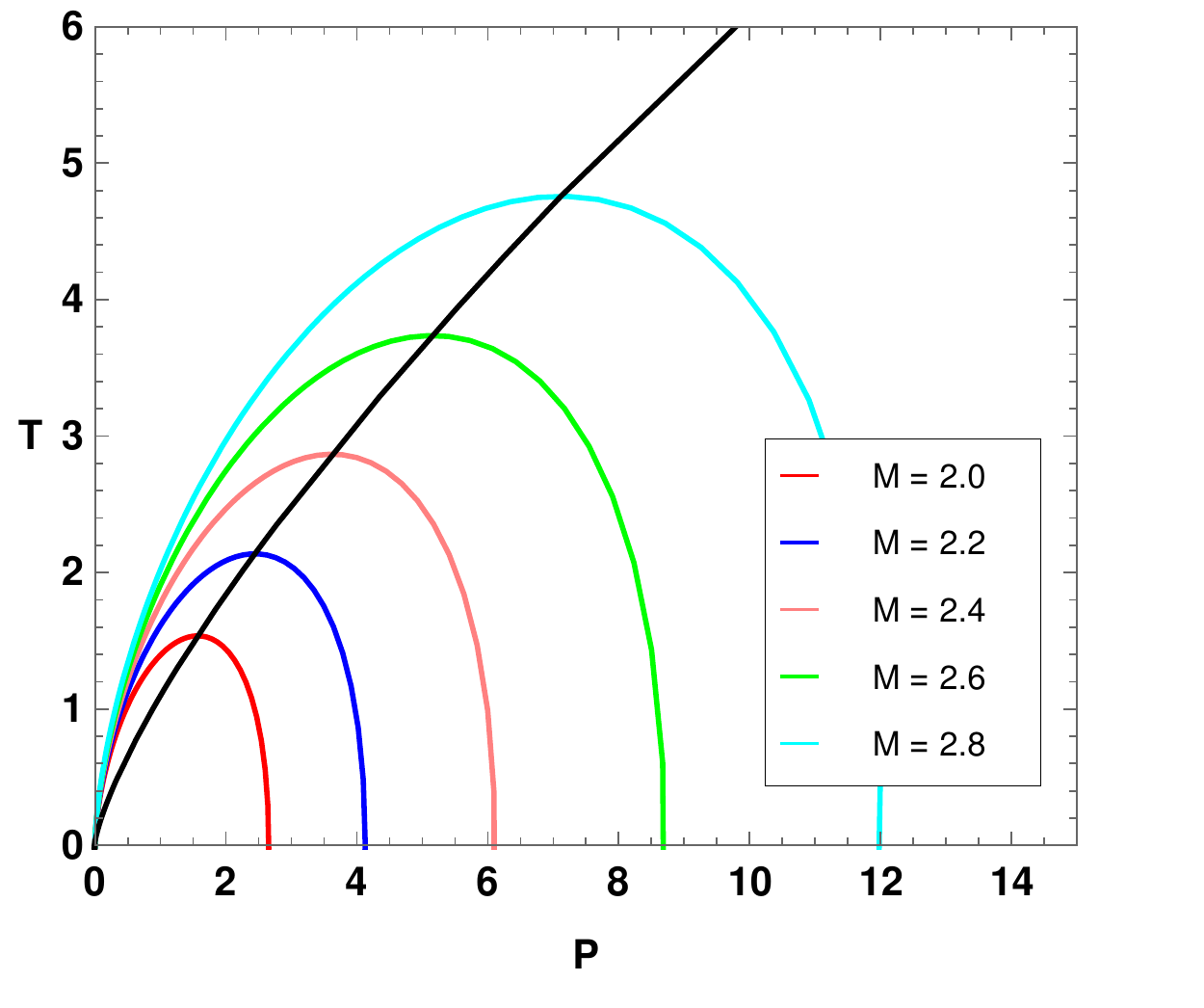}
       }
      	\caption{Isenthalpic and inversion curves of the black hole defined by Eq. \eqref{metric01}. The black line represents the inversion curve. Coloured curved lines represent the isenthalpic curves corresponding to different values of black hole mass $M$. Here, we have used $N_s = 0.1$, $\beta = 0.1$ and $Q=1.5$ (on first panel) and $Q=0.9$ (on second panel).}
      	\label{figJT01}
      \end{figure}

\begin{figure}[h]
      	\centering{
      		\includegraphics[scale=0.55]{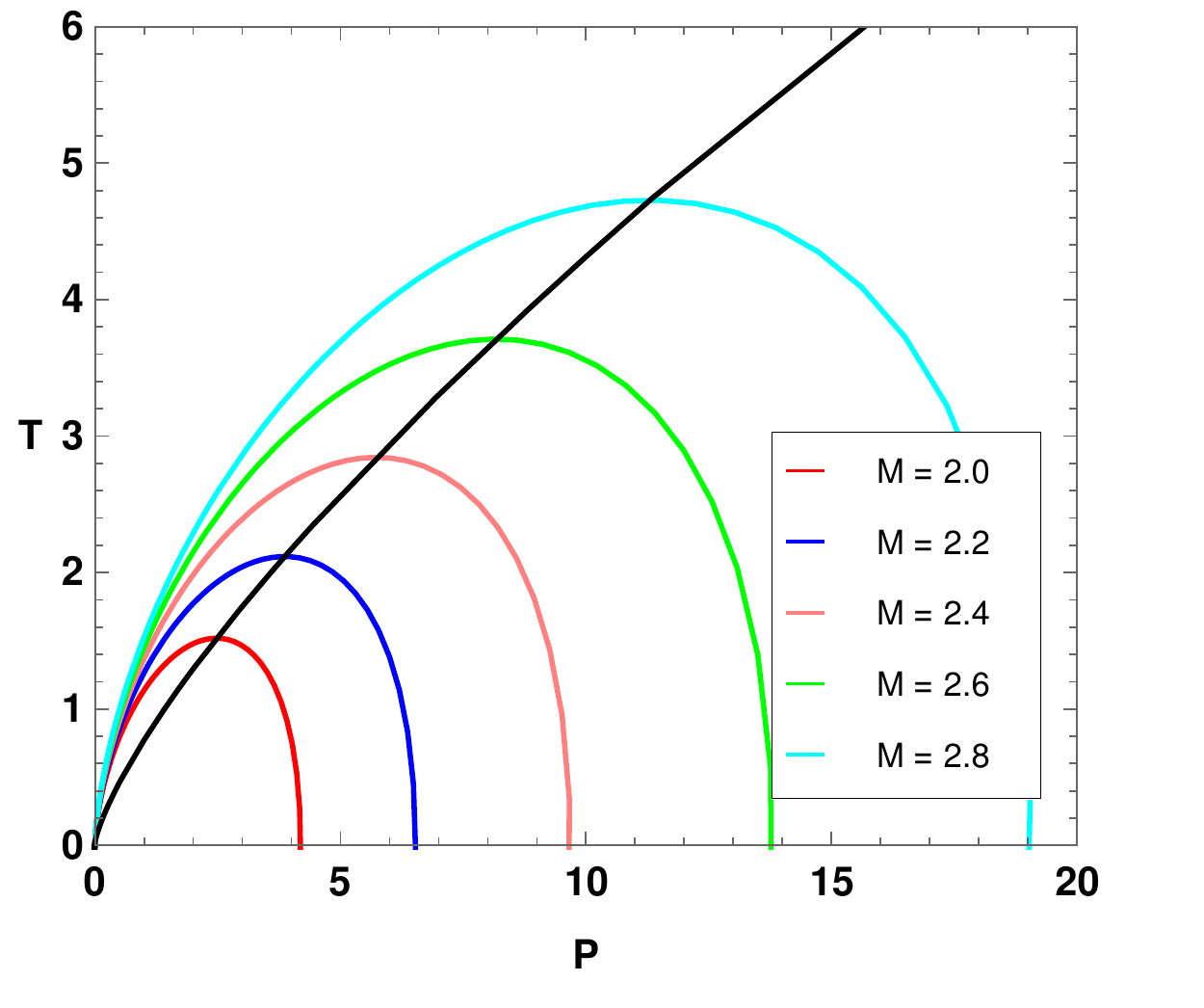}
       \includegraphics[scale=0.55]{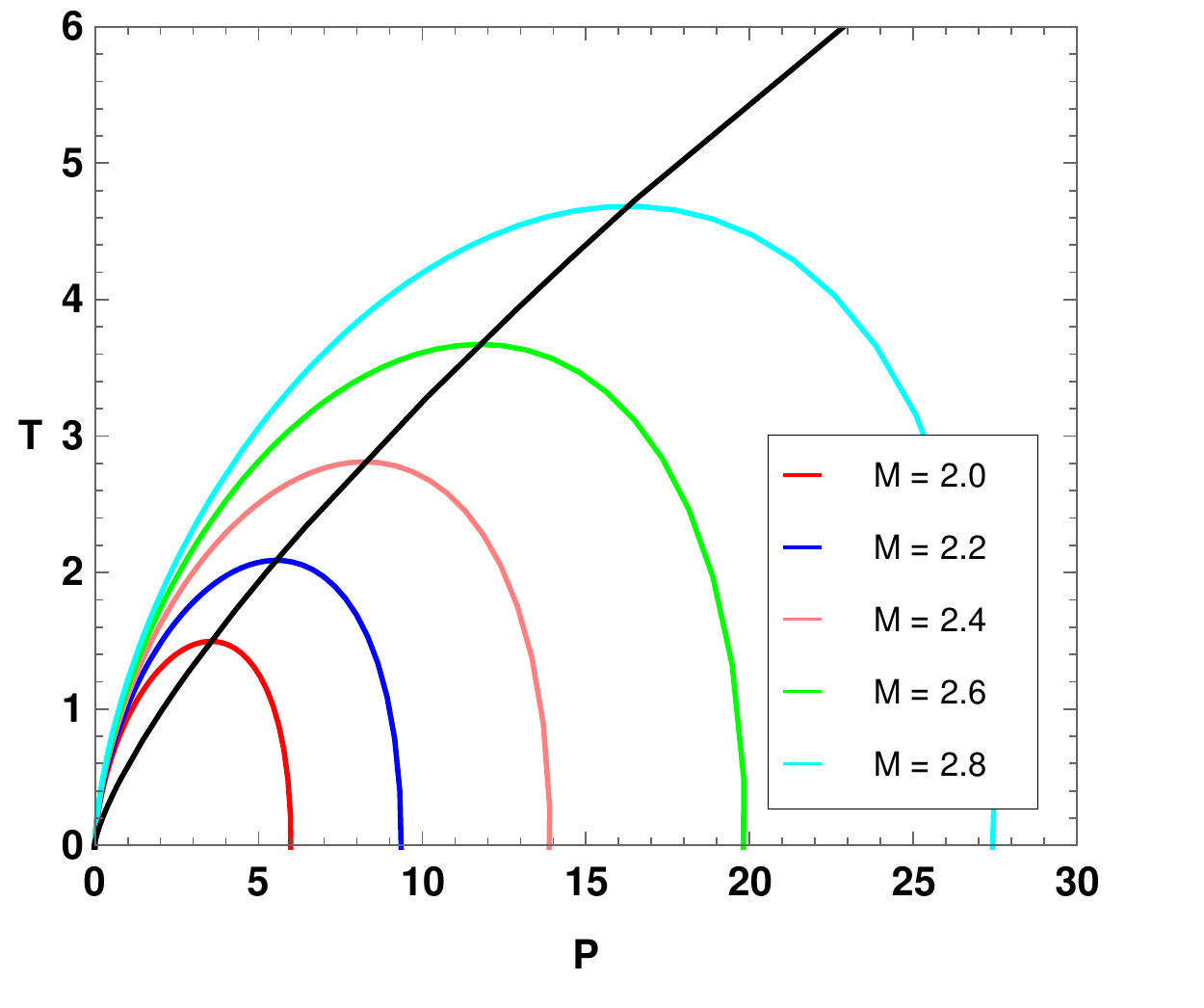}
       }
      	\caption{Isenthalpic and inversion curves of the black hole defined by Eq. \eqref{metric01}. The black line represents the inversion curve. Coloured curved lines represent the isenthalpic curves corresponding to different values of black hole mass $M$. Here, we have used $Q=0.9$, $N_s = 0.1$ and $\beta = 0.1$ (on first panel) and $\beta = -0.1$ (on second panel).}
      	\label{figJT02}
      \end{figure}

\begin{figure}[h]
      	\centering{
      		\includegraphics[scale=0.55]{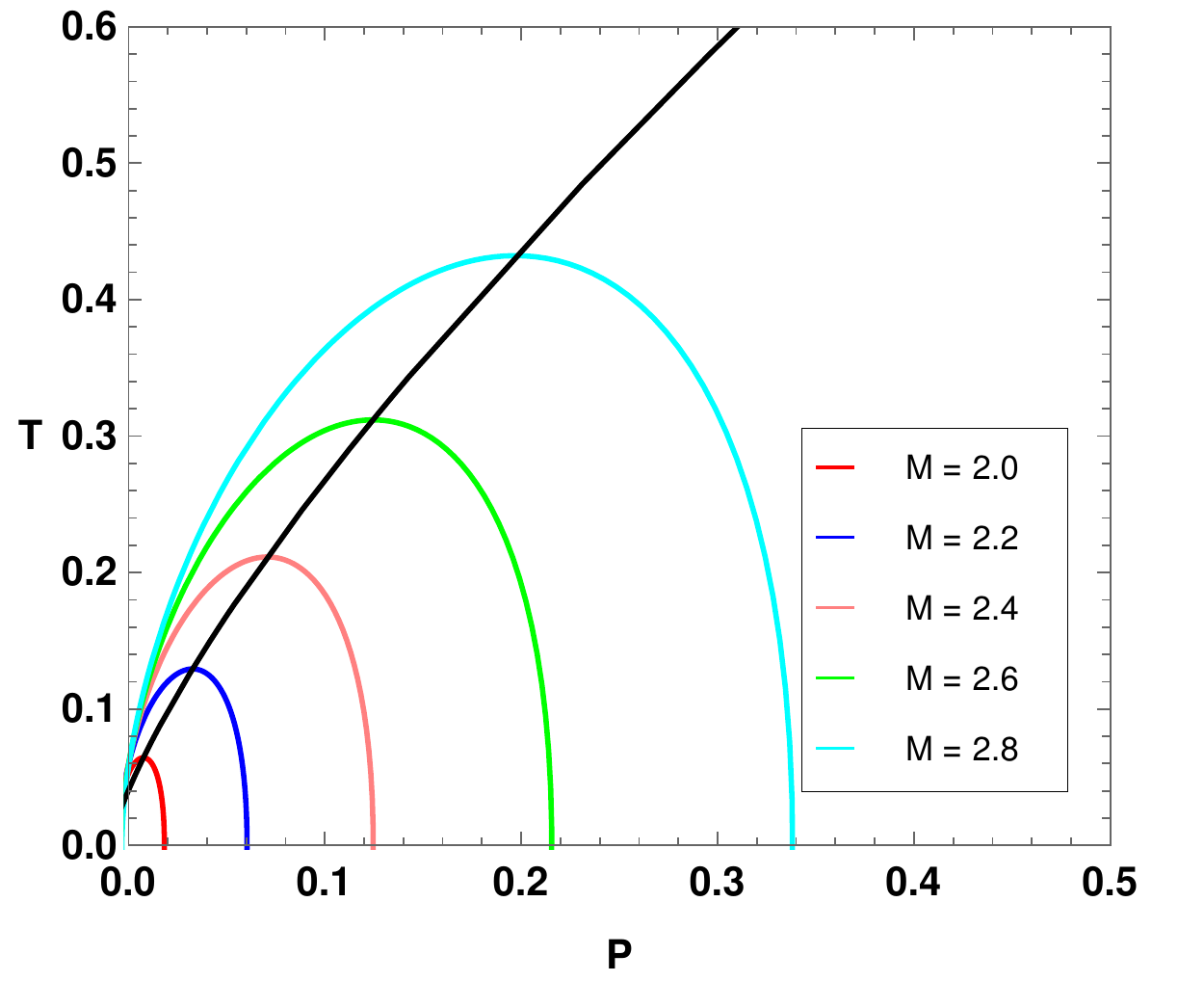}
       }
      	\caption{Isenthalpic and inversion curves of the black hole defined by Eq. \eqref{metric01}. The black line represents the inversion curve. Coloured curved lines represent the isenthalpic curves corresponding to different values of black hole mass $M$. Here, we have used $Q=1.5$, $N_s = 0.5$ and $\beta = 0.1$.}
      	\label{figJT03}
      \end{figure}

Finally, we have shown the variation of the isenthalpic curves and inversion curves of the black hole in Fig.s \ref{figJT01}, \ref{figJT02} and \ref{figJT03} for different sets of model parameters. One can see that in these figures, the isenthalpic curves are intersected by the inversion curves into two different parts. The upper part basically depicts the cooling region of the black hole, while the lower part of the curves denotes the heating phase. 

From Fig. \ref{figJT01}, we see that the inversion temperature increases with an increase in the mass of the black hole. However, with a decrease in the charge $Q$ of the black hole, the inversion temperature increases significantly.

In Fig. \ref{figJT02}, one can see that with an increase in the Rastall parameter, the inversion temperature also increases. In the figure, the inversion temperature for the case with $\beta = 0.1$ is slightly higher. However, a significant change in the corresponding inversion pressure $P_i$ is seen. The case with $\beta = -0.1$ has higher pressures corresponding to the inversion temperature $T_i$. Such a variation has significant impacts on the cooling and heating phase of the black hole, as depicted in the figure. 

In Fig. \ref{figJT03}, we have used $Q=1.5$, $\beta = 0.1$ and $N_s = 0.5$ to depict the isenthalpic and inversion curves. Comparing this figure with the first panel of Fig. \ref{figJT01}, we can observe that an increase in $N_s$ decreases the inversion temperature $T_i$. An increase in the structural parameter $N_s$, hence decreases both the cooling and heating phase or region of the black hole.
 
\section{Optical Properties of the black hole}\label{sec06}
\subsection{Shadow behaviors}
Showing the optical behaviour pertinent to our black hole solution is the primary objective of this section. Well, taking advantage of many works regarding shadow behaviours \cite{CS,AB} helps us discover in an easy way a one-dimensional closed curve on the image plane. Indeed, an examination of the optical aspect, particularly the shadow image, arises from considering the Hamilton-Jacobi action \cite{BC,YD}
\begin{equation}
    \frac{\partial \mathcal{S}}{\partial \delta}+H=0,
    \label{E5-1}
\end{equation}
where $\mathcal{S}$ and $\delta$ are the Jacobi action and the affine parameter along the geodesics, respectively. A concrete visualisation of the shadow image in the static, spherically symmetric spacetime is carried out with a massless photon described by the following Hamiltonian
\begin{equation}
    H=\frac{1}{2}g^{ij}p_ip_j=0.
    \label{E5-2}
\end{equation}
In fact, the spherically symmetric feature of the black hole causes the photon motion to be on the equatorial plane $\theta=\frac{\pi}{2}$. As a consequence, the Eq.(\ref{E5-2}) gives
\begin{equation}
\frac{1}{2}\left[-r^2f(r)p_t^2+\left(rf(r)p_r\right)^2+f(r)p_\phi^2\right]=0.
    \label{E5-3}
\end{equation}
Here, $E=-p_t$ and $L=p_{\phi}$ are the conserved quantities matching, respectively, the total energy and the angular momentum of the photon. Using the Hamiltonian formalism, the compact expressions of the equations of motion are given by
\begin{eqnarray}
\frac{dt}{d\delta}&=&\frac{\partial H}{\partial p_t}=-\frac{p_t}{f(r)},\\
\frac{dr}{d\delta}&=&\frac{\partial H}{\partial p_r}=p_r f(r),\\
\frac{d\phi}{d\delta}&=&\frac{\partial H}{\partial p_\phi}=\frac{p_\phi}{r^2}
\label{E5-4}
\end{eqnarray}
To have a better understanding of the corresponding shadow behaviours, we must first establish an effective potential. This can show that the shape of a black hole is entirely defined by the boundaries of its shadow, which is the apparent shape of the photons' unstable circular orbits. The pursuit of this objective requires presenting the effective potential in such a way as
\begin{equation}
    \left(\frac{dr}{d\delta}\right)^2+V_{eff}(r)=0.
    \label{E5-5}
\end{equation}
In other terms, the previous expression provides the equivalent of the following:
\begin{equation}
    V_{eff}(r)=f(r)\left[\frac{L^2}{r^2}-\frac{E^2}{f(r)}\right].
    \label{E5-6}
\end{equation}
To indicate the circular orbit as well as the unstable photons, all of these are acquired by the given maximum value of the effective potential through the following constraints:
\begin{equation}
    V_{eff}=\frac{dV_{eff}(r)}{dr}\biggr\rvert_{r=r_p}=0.
    \label{E5-7}
\end{equation}
By the way, applying these constraints yields
\begin{eqnarray}
f(r)\left(\frac{L^2}{r_p^2}-\frac{E^2}{f(r_p)}\right)&=&0,\\
\frac{L^2}{r_p^2}\cdot\frac{r_pf'(r_p)-2f(r_p)}{r_p}&=&0.
\label{E5-8}
\end{eqnarray}
It is noted that $f'(r)$ is a notation of the differentiation $\frac{\partial f}{\partial r}$. With $r_p$ is the photon sphere radius of matter and gravity background in $AdS$ space and represents a solution of the following equation:
\begin{equation}
    r_pf'(r_p)-2f(r_p)=0.
    \label{E5-9}
\end{equation}
It emerges from Eq.($\ref{E5-9}$) that finding $r_p$ analytically is difficult. Alternatively, we can solve the related equation using a numerical approach. The parametric orbit equation for photon motion is clearly given based on consideration of the equation
\begin{equation}
    \frac{dr}{d\phi}=\pm r\sqrt{f(r)\left(\frac{r^2 E^2}{f(r)L^2}-1\right)}.
    \label{E5-10}
\end{equation}
It should be emphasized that the turning point of the photon orbit is constrained by
\begin{equation}
    \frac{dr}{d\phi}\biggr\rvert_{r=R}=0.\label{E5-11}
\end{equation}
Hence, the latter equation provides
\begin{equation}
     \frac{dr}{d\phi}=\pm r\sqrt{f(r)\left(\frac{r^2 f(R)}{R^2 f(r)}-1\right)}.
     \label{E5-12}
\end{equation}
To generate a nice description of the shadow image, consider a light ray originating at $r_O$ and propagating at an angle into the past with an angle $\Theta$ with respect to the radial direction. Thus, one can obtain \cite{ZG,CM}
\begin{equation}
    \cos\Theta=\sin\Theta\left(\frac{\sqrt{g_{rr}}}{\sqrt{g_{\phi\phi}}}\right)\frac{dr}{d\phi}\biggr\rvert_{r=r_O}.
     \label{E5-13}
\end{equation}
As a result, the shadow radius of the black hole is completely specified as
\begin{equation}
    r_s=r_O\sin\Theta=R\sqrt{\frac{f(r_O)}{f(R)}}\biggr\rvert_{R=r_p}.
    \label{E5-14}
\end{equation} 
A static observer at spatial infinity has $f(r_O)=1$ \cite{ZG}. The apparent shape of the shadow is computed by implementing the celestial coordinates $X$, and $Y$, which are expressed as \cite{SV}
\begin{eqnarray}
X&=&\underset{r_O\rightarrow\infty}{lim}\left(-r_O^2\sin\theta_O\frac{d\phi}{dr}\biggr\rvert_{(r_O,\theta_O)}\right)\\
Y&=&\underset{r_O\rightarrow\infty}{lim}\left(r_O^2\frac{d\theta}{dr}\biggr\rvert_{(r_O,\theta_O)}\right)
\label{E5-15}
\end{eqnarray}

\begin{figure}[htb!]%
    \centering
    \subfloat{{\includegraphics[width=7cm]{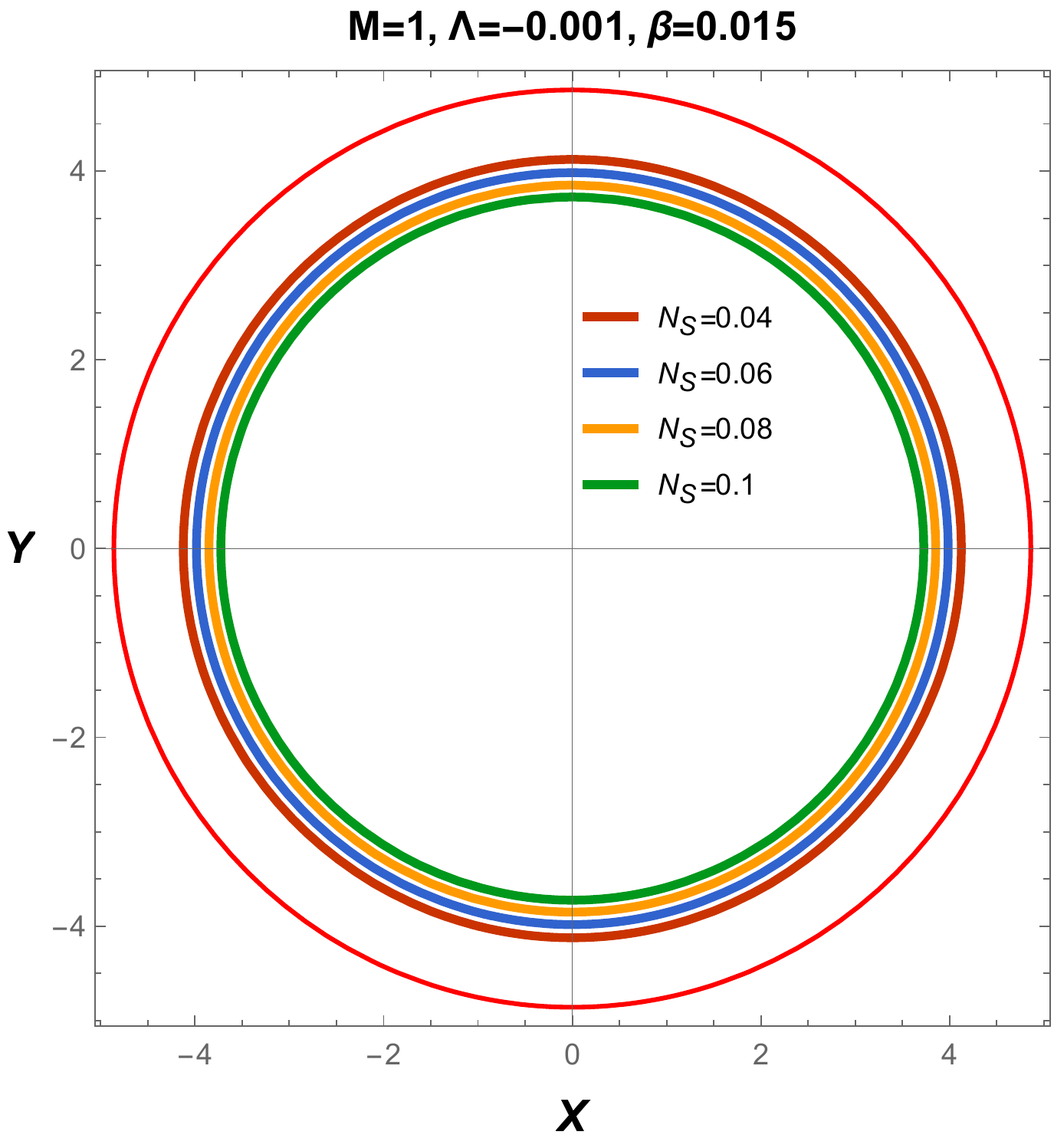} }}%
    \qquad
    \subfloat{{\includegraphics[width=7cm]{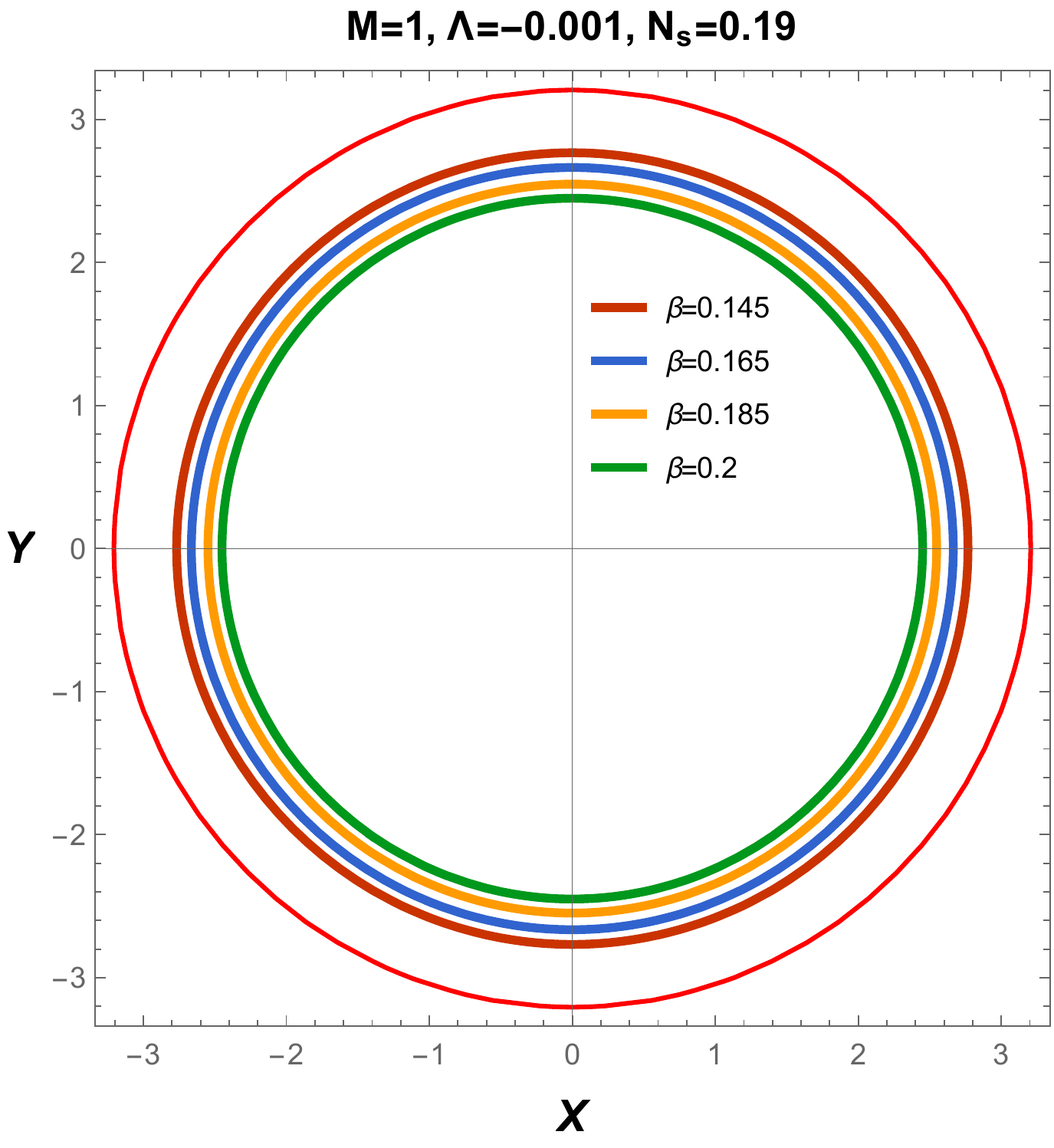} }}%
    \caption{Black hole shadow in the celestial plane ($X$-$Y$) for different values of the black hole structural parameter $N_s$ and the Rastall parameter $\beta$. The red circle represents the uncharged case $(Q=0)$, the upper panel $N_s=0.04$, and the lower panel $\beta=0.145$.}%
    \label{fig1}%
\end{figure}
\begin{figure}[hbt!]
      	\centering
      	\includegraphics[scale=0.45]{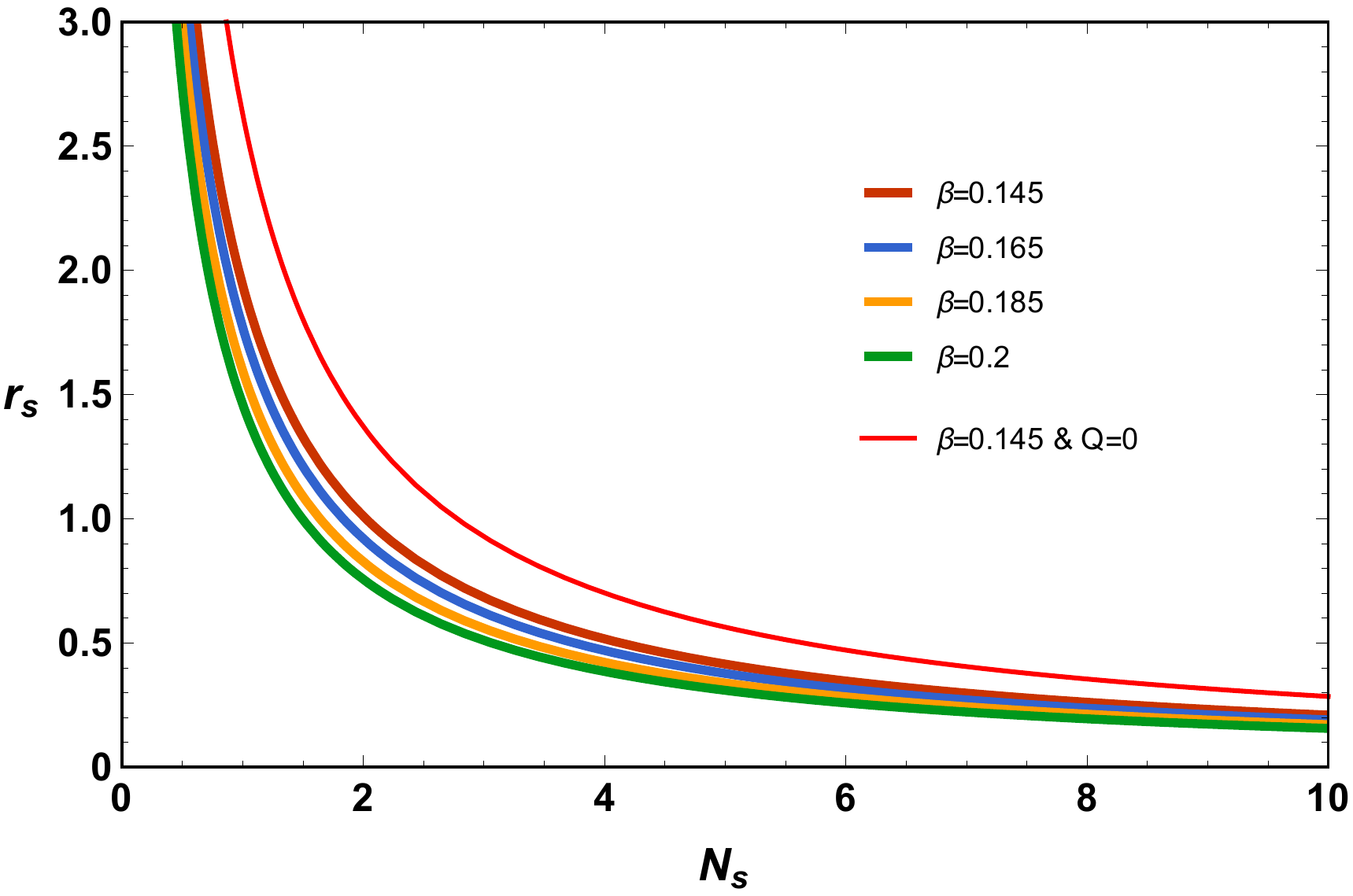}
      	\caption{The variation of the shadow radius in terms of the black hole structural parameter $N_s$ for different fixed values of the Rastall parameter $\beta$ for $Q=0.85$ and $\Lambda=-0.01$.}
      	\label{fig2}
      \end{figure}

where $(r_0, \theta_0)$ are the coordinates of the observer. On the other hand, the numerical approach aims to draw conclusions from what we have done using the analytical method. For that reason, we collect all of the physical parameters describing the black hole system to construct a parameter space labelled $\mathcal{M}\left(M, Q,\Lambda, N_s, \beta\right)$. $M$, like the black hole mass, is fixed at $1$. Moreover, among the different possible values of the electric charge, we set $Q = 0.85$ along with the cosmological constant $\Lambda = -0.001$. Thus, the remaining parameter space is now spanned out via its partial version, which $\mathcal{M}\left(N_s, \beta\right)$. Taking these fixed values at hand, we depict with respect to several values of $N_s$ and $\beta$ the shadow behaviours related to the black hole solution. Fig. \ref{fig1} shows the variation of the shadow radius as a number of $one$-$dimensional$ closed curves with respect to different values of $N_s$ and $\beta$. It results from the two first figures that the shadow size is 
 inversely proportional to both $N_s$ and $\beta$; clearly, when the $N_s$ and $\beta$ values increase, the size of the shadow decreases. Furthermore, a comparison of the charged versus uncharged cases reveals that once the electric charge is devoid of spacetime, the shadow size rapidly increases $\left(r_s^2\rvert_{Q=0}>r_s^2\rvert_{Q\neq0}\right)$. To see how the shadow radius varies with parameters such as the black hole structural parameter, $N_s$, Fig. \ref{fig2} shows this dependence in such a way that the shadow radius is a decreasing function of the black hole structural for a fixed photon sphere radius.
\subsection{Energy emission rate}
This part takes care of studying the relevant energy emission rate. Due to the process causing the formation and annihilation of certain pairs of particles near the horizons of the black hole, particles of positive energy can escape from the black hole through tunnels within the region where Hawking radiation arises. This is referred to as Hawking radiation, and it leads the black hole to evaporate over time. It was demonstrated that the absorption cross-section approaches the black hole shadow \cite{WL} for a faraway observer. At very high energies, the absorption cross-section behaves as an oscillator with a limiting constant value of ($\sigma_{lim}=\pi r_s^2$). Remarkably, the corresponding energy emission rate can be expressed in the following way:
\begin{equation}
    \frac{d^2E(\varpi)}{dtd\varpi}=\frac{2\pi^3\varpi^3 r_s^2}{e^{\frac{\varpi}{T}}-1},
    \label{E5-16}
\end{equation}
where $\varpi$ denotes the emission frequency and $T$ the Hawking temperature. 
\begin{figure}[t!]%
    \centering
    \subfloat{{\includegraphics[width=8cm]{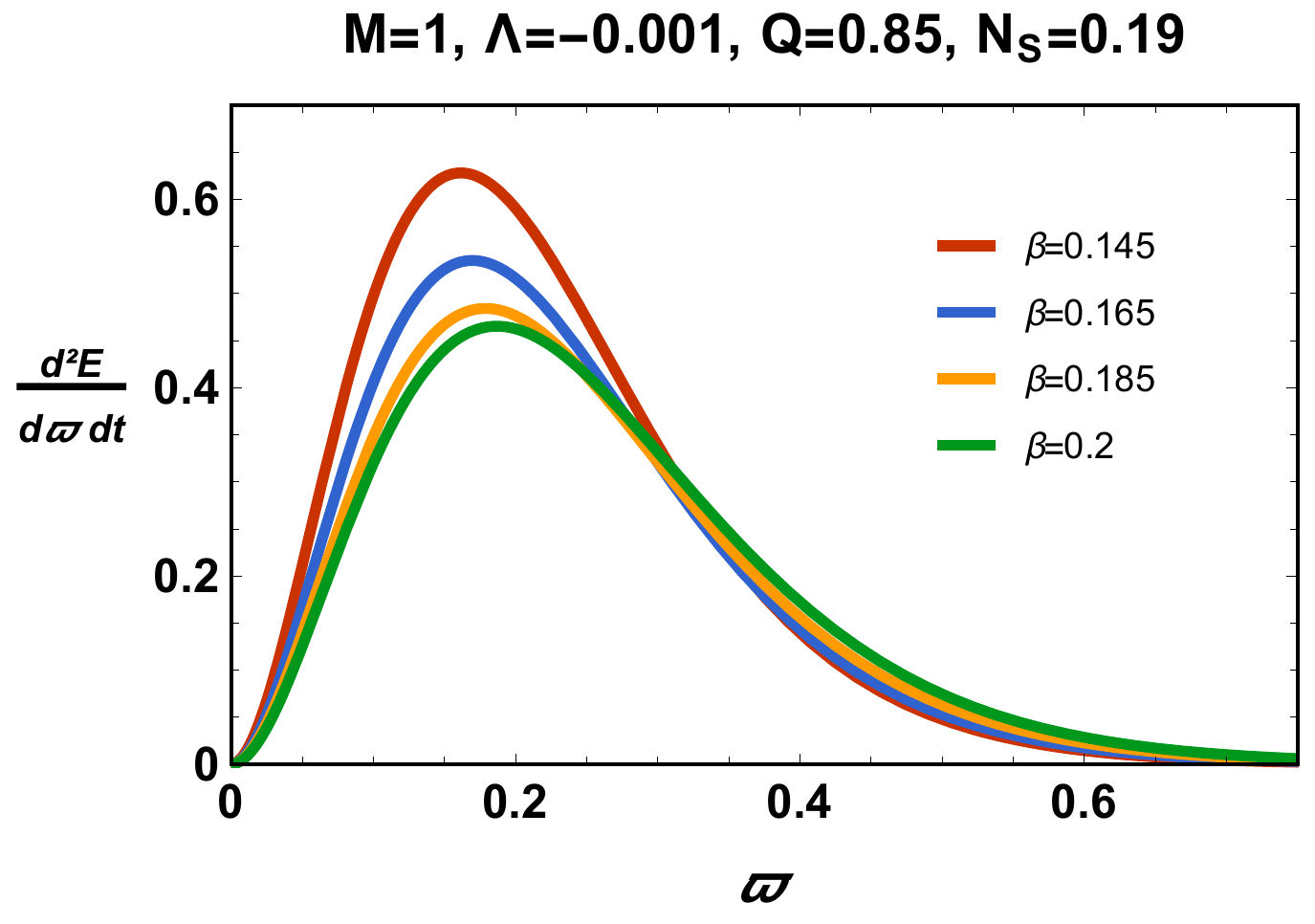} }}%
    \qquad
    \subfloat{{\includegraphics[width=8cm]{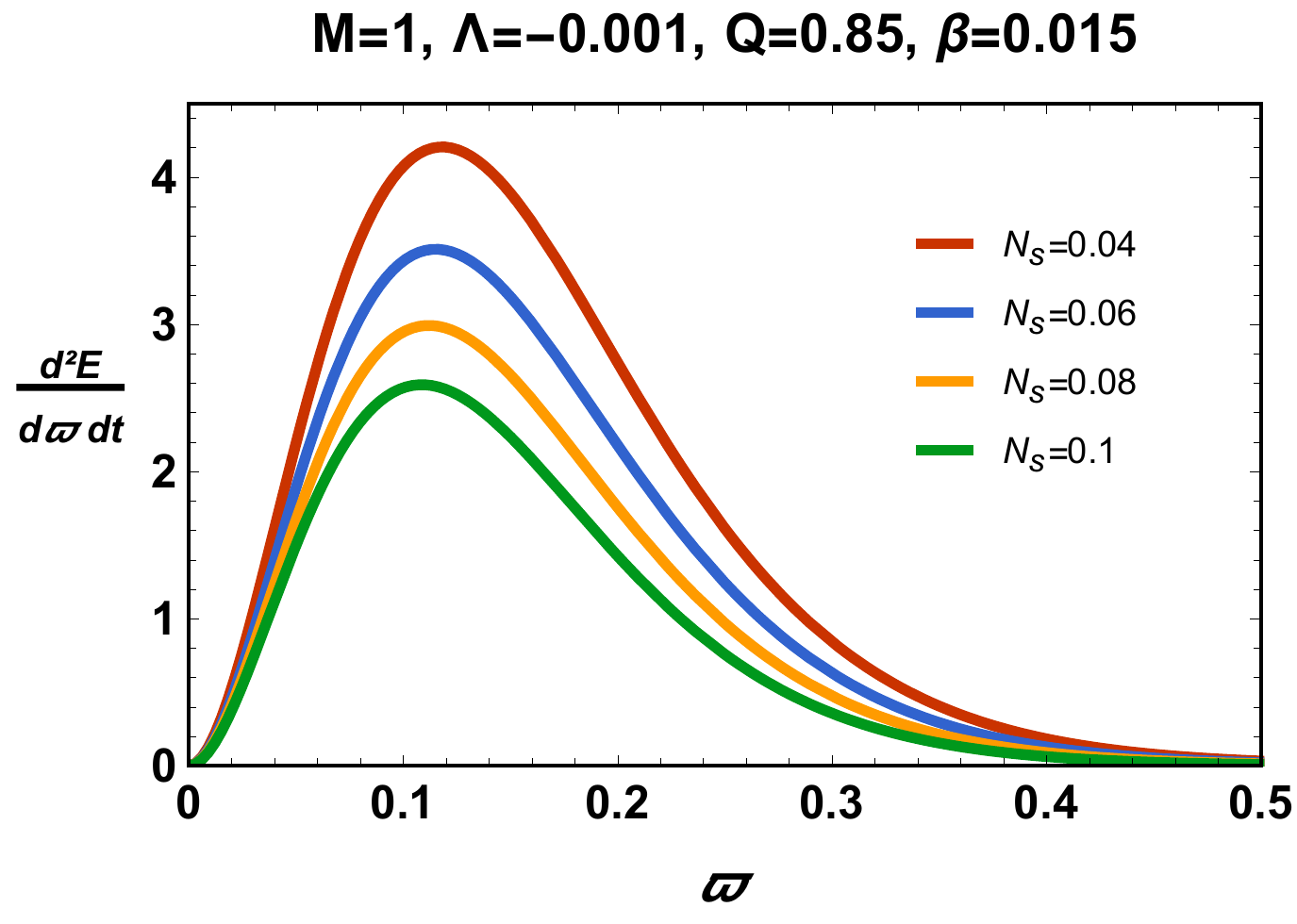} }}%
    \caption{The energy emission rate of the black hole for different values of the black hole structural parameter $N_s$ and the Rastall parameter $\beta$.}%
    \label{fig3}%
\end{figure}
To inspect the energy emission rate behaviour graphically, Fig. \ref{fig3} shows the variation of the energy emission rate as a function of $\varpi$ for various fixed values of $N_s$ and $\beta$. According to Fig. \ref{fig3}, both the $N_s$ and $\beta$ parameter variations have the same effect on the energy emission rate. The observation reveals that with high fixed values of the parameters $N_s$ and $\beta$, the energy emission rate decreases significantly. This means that the black hole evaporation process is slow.

\section{Conclusion}\label{sec07}
In this work, we have studied thermodynamics, Joule-Thomson expansion and optical behaviour of a Reissner-Nordstr\"om-anti-de Sitter black hole in Rastall gravity surrounded by a quintessence field. The black hole solution is singular in nature and significantly different from the analogous GR black hole. We observe that the Rastall parameter $\beta$ can impact the Joule-Thomson expansion of the black hole. Although this impact is smaller in comparison, the investigation depicted that with an increase in $\beta$, the inversion temperature increases slightly, and the corresponding pressure decreases significantly. The structural parameter $N_s$ decreases the inversion temperature. The heating and cooling phases are also impacted significantly by these parameters. The Joule-Thomson coefficient is highly impacted by the black hole charge, and the impact is almost negligible for a slight variation of the parameter $\beta$.

The analysis also depicts that the black hole shadow and the energy emission rate are impacted by the Rastall parameter $\beta$ and the black hole structural parameter $N_s$. An increase in the black hole structural parameter $N_s$ results in a decrease in the shadow radius. The presence of a violation of energy-momentum conservation also decreases the shadow radius. The energy emission rate of the black hole is higher for smaller values of the black hole structural parameter $N_s$, and with an increase in the Rastall parameter $\beta$, the energy emission rate decreases gradually. Hence, the investigation depicts that the black hole evaporation rate decreases with an increase in both $N_s$ and $\beta$. It implies that in Rastall gravity, a black hole may have a higher lifetime with slow evaporation.

\bibliographystyle{apsrev}

\begin{thebibliography}{99}

\bibitem{2016_Abbott}
B. P. Abbott et al., {\em Observation of Gravitational Waves from a Binary Black Hole Merger}, 
{Phys. Rev. Lett. {\bf 116}, 061102 (2016)}.

\bibitem{will2014}
C. M. Will, {\em The Confrontation between General Relativity and Experiment}, 
{Living Rev. Relativ. {\bf 17}, 4 (2014)}.

\bibitem{hulse1975}
R. A. Hulse and J. H. Taylor, {\em Discovery of a Pulsar in a Binary System}, {ApJL {\bf 195}, L51 (1975)}.

\bibitem{damour1992}
T. Damour and J. H. Taylor, {\em Strong-Field Tests of Relativistic Gravity and Binary Pulsars}, {Phys. Rev. D {\bf 45}, 1840 (1992)}.

\bibitem{Liang_2017}
D. Liang, Y. Gong, S. Hou and Y. Liu, {\em Polarizations of gravitational waves in f(R) gravity}, 
{Phys. Rev. D {\bf 95}, 104034 (2017)}.

\bibitem{gogoi1}
D. J. Gogoi and U. D. Goswami, {\em A New f(R) Gravity Model and Properties of Gravitational Waves in It}, {Eur. Phys. J. C {\bf 80}, 1101 (2020)}.

\bibitem{gogoi2}
D. J. Gogoi and U. Dev Goswami, {\em Gravitational Waves in $\mathbf {f(R)}$ Gravity Power Law Model}, Indian J. Phys. {\bf 96}, 637 (2022).

\bibitem{1972_Rastall}
P. Rastall, {\em Generalization of the Einstein Theory}, {Phys. Rev. D {\bf 6}, 3357 (1972)}.

\bibitem{Ferrari}
V. Ferrari and L. Gualtieri, {\em Quasi-Normal Modes and Gravitational Wave Astronomy}, {Gen. Relativ. Gravit. {\bf 40}, 945 (2008)}.

\bibitem{Oliveira}
A. M. Oliveira, H. E. S. Velten, J. C. Fabris and L. Casarini, {\em Neutron stars in Rastall gravity}, 
{Phys. Rev. D {\bf 92}, 044020 (2015)}.

\bibitem{Heydarzade}
Y. Heydarzade, H. Moradpour, and F. Darabi, {\em Black Hole Solutions in Rastall Theory}, Can. J. Phys. {\bf 95}, 1253 (2017).

\bibitem{Heydarzade2}
Y. Heydarzade and F. Darabi, {\em Black Hole Solutions Surrounded by Perfect Fluid in Rastall Theory}, Phys. Lett. B {\bf 771}, 365 (2017).

\bibitem{Xu}
Z. Xu, X. Hou, X. Gong, and J. Wang, {\em Kerr-Newman-AdS Black Hole Surrounded by Perfect Fluid Matter in Rastall Gravity}, Eur. Phys. J. C {\bf 78}, 513 (2018).

\bibitem{Lin}
K. Lin and W.-L. Qian, {\em Neutral Regular Black Hole Solution in Generalized Rastall Gravity}, Chinese Phys. C {\bf 43}, 083106 (2019).

\bibitem{Hu}
Y. Hu, C.-Y. Shao, Y.-J. Tan, C.-G. Shao, K. Lin, and W.-L. Qian, {\em Scalar Quasinormal Modes of Nonlinear Charged Black Holes in Rastall Gravity}, EPL {\bf 128}, 50006 (2020).

\bibitem{Liang}
J. Liang, {\em Quasinormal Modes of the Schwarzschild Black Hole Surrounded by the Quintessence Field in Rastall Gravity}, Commun. Theor. Phys. {\bf 70}, 695 (2018).

\bibitem{gogoi3}
D. J. Gogoi and U. D. Goswami, {\em Quasinormal Modes of Black Holes with Non-Linear-Electrodynamic Sources in Rastall Gravity}, Physics of the Dark Universe {\bf 33}, 100860 (2021).

\bibitem{gogoi4}
D. J. Gogoi, R. Karmakar, and U. D. Goswami, {\em Quasinormal Modes of Nonlinearly Charged Black Holes Surrounded by a Cloud of Strings in Rastall Gravity}, Int. J. Geom. Methods Mod. Phys. {\bf 20}, 2350007 (2023).

\bibitem{mu}
\"O. \"Okc\"u and E. Aydıner, {\em Joule–Thomson Expansion of the Charged AdS Black Holes}, Eur. Phys. J. C {\bf 77}, 24 (2017).

\bibitem{Meng2020}
Y. Meng, J. Pu, and Q.-Q. Jiang, {\em P-V Criticality and Joule-Thomson Expansion of Charged AdS Black Holes in the Rastall Gravity}, Chinese Phys. C {\bf 44}, 065105 (2020).

\bibitem{Cao2021}
Y. Cao, H. Feng, W. Hong, and J. Tao, {\em Joule-Thomson Expansion of RN-AdS Black Hole Immersed in Perfect Fluid Dark Matter}, Commun. Theor. Phys. {\bf 73}, 095403 (2021).

\bibitem{Yin2021}
R. Yin, J. Liang, and B. Mu, {\em Joule–Thomson Expansion of Reissner–Nordstr\"om-Anti-de Sitter Black Holes with Cloud of Strings and Quintessence}, Physics of the Dark Universe {\bf 34}, 100884 (2021).

\bibitem{jt01}
A. Biswas, {\em Joule-Thomson Expansion of AdS Black Holes in Einstein Power-Yang-Mills Gravity}, Phys. Scr. {\bf 96}, 125310 (2021).

\bibitem{jt02}
Y. Cao, H. Feng, J. Tao, and Y. Xue, {\em Black Holes in a Cavity: Heat Engine and Joule-Thomson Expansion}, Gen. Relativ. Gravit. {\bf 54}, 105 (2022).

\bibitem{jt03}
J. P. Morais Graça, E. F. Capossoli, and H. Boschi-Filho, {\em Joule-Thomson Expansion for Noncommutative Uncharged Black Holes}, EPL {\bf 135}, 41002 (2021).

\bibitem{jt04}
J. Liang, B. Mu, and P. Wang, {\em Joule-Thomson Expansion of Lower-Dimensional Black Holes}, Phys. Rev. D {\bf 104}, 124003 (2021).

\bibitem{jt05}Y. Sekhmani, Z. Dahbi, A. Najim, and A. Waqdim, {\em Joule-Thomson Expansion of 5-Dimensional R-Charged Black Holes}, Annals of Physics {\bf 444}, 169060 (2022).

\bibitem{Balart}
L. Balart and E. C. Vagenas, Regular Black Holes with a Nonlinear Electrodynamics Source, Phys. Rev. D {\bf 90}, 124045 (2014).

\bibitem{Kiselev}
V. V. Kiselev, \textit{Quintessence and Black Holes}, Class. Quantum. Grav {\bf 20}, 1187 (2003) {\tt [arXiv:gr-qc/0210040]}.

\bibitem{CS}
A.~Belhaj and Y.~Sekhmani, {\em Shadows of rotating quintessential black holes in Einstein\textendash{}Gauss\textendash{}Bonnet gravity with a cloud of strings}, 
{Gen. Rel. Grav. \textbf{54}, no.2, 17 (2022)}.

\bibitem{AB}A.~Belhaj and Y.~Sekhmani,
\textit{Optical and thermodynamic behaviors of Ay\'on\textendash{}Beato\textendash{}Garc\'\i{}a black holes for 4D Einstein Gauss\textendash{}Bonnet gravity},
{Annals Phys. \textbf{441}, 168863 (2022)}.

 \bibitem{BC}
B. Carter, \textit{Global structure of the Kerr family of gravitational fields}, Phys. Rev. \textbf{174}, 1559 (1968).

\bibitem{YD} Y. D{\'e}canini, A. Folacci, and R. Bernard. \textit{Fine structure of high-energy absorption cross-sections for black holes}, Class. Quantum Grav. \textbf{28} 175021 (2011).

\bibitem{ZG}M. Zhang and M. Guo, \textit{Can shadows reflect phase structures of black holes?}, Eur. Phys. J. C \textbf{80} 8, 790 (2020) {\tt arXiv:1909.07033 [gr-qc].}

\bibitem{CM}A. Belhaj, L. Chakhchi, H. El. Moumni, et al., \textit{Thermal image and phase transitions of charged AdS black holes using shadow analysis}, Int. J. Mod. Phys. A. \textbf{35} 2050170 (2020).

\bibitem{SV}S. E Vazquez and E. P Esteban, \textit{Strong field gravitational lensing by a Kerr black hole}, Nuovo Cim.B \textbf{119} 489 (2004).

\bibitem{WL}S. W. Wei and Y. X. Liu, \textit{Observing the shadow of Einstein-Maxwell-Dilaton-Axion black hole}, JCAP \textbf{11} 063 (2013), {\tt arXiv:1311.4251 [gr-qc].}

\end{thebibliography}
\end{document}